\pdfoutput=1
\documentclass[twocolumn,prb,amsmath,superscriptaddress,
               preprintnumbers,amssymb,amsfonts,showpacs,
               letterpaper,floatfix]{revtex4}
\usepackage{graphicx} 
\usepackage{epsfig}

\begin{document}


\title{Electronic properties of bulk and thin film SrRuO$_3$: a search\\ 
       for the metal-insulator transition}
  \author{James M.\ Rondinelli}
  \affiliation{Materials Department, University of California, Santa Barbara, 
	       CA, 93106-5050, USA}
  \author{Nuala M.\ Caffrey}
  \affiliation{School of Physics and CRANN, Trinity College, Dublin 2, Ireland}
  \author{Stefano Sanvito}
  \affiliation{School of Physics and CRANN, Trinity College, Dublin 2, Ireland}
  \author{Nicola A.\ Spaldin}
     \email[Address correspondence to: ]{nicola@mrl.ucsb.edu}
  \affiliation{Materials Department, University of California, Santa Barbara, 
	       CA, 93106-5050, USA}
\date{\today}

\begin{abstract}
We calculate the properties of the 4$d$ ferromagnet SrRuO$_3$ in bulk and 
thin film form with the aim of understanding the experimentally observed 
metal to insulator transition at reduced thickness. 
Although the spatial extent of the 4$d$ orbitals is quite large, many 
experimental results have suggested that electron-electron 
correlations play an important role in determining this material's electronic
structure.
In order to investigate the importance of correlation, we use two approaches 
which go beyond the conventional local density approximation to density 
functional theory (DFT):
the local spin density approximation + Hubbard $U$ (LSDA+$U$) and the 
pseudopotential self-interaction correction (pseudo-SIC) methods.
We find that the details of the electronic structure predicted with the LSDA 
do not agree with the experimental spectroscopic data for bulk and thin film 
SrRuO$_3$.
Improvement is found by including electron-electron correlations, and we suggest 
that bulk orthorhombic SrRuO$_3$ is a {\it weakly strongly-correlated} ferromagnet 
whose electronic structure is best described by a 0.6~eV on-site Hubbard 
term, or equivalently with corrections for the self-interaction error.
We also perform {\it ab initio} transport calculations that confirm that SrRuO$_3$ 
has a negative spin polarization at the Fermi level, due to the position of the 
minority Ru 4$d$ band center.
Even with correlations included in our calculations we are unable to reproduce
the experimentally observed metal-insulator transition, suggesting that the
electronic behavior of SrRuO$_3$ ultra-thin films might be dominated by extrinsic
factors such as surface disorder and defects. 
\end{abstract}

\maketitle

\section{Introduction}
Conductive oxides are essential components in composite oxide 
heterostructures where they are often used as electrode materials in thin film 
applications.\cite{Marrec_et_al:2002,Takahashi/Tokura_et_al:2005,Gallagher/Parkin_2006}
In the perovskite crystal family ($AB$O$_3$ stoichiometry), the itinerant 
ferromagnetic SrRuO$_3$ (SRO) is a popular choice since it is one of the more 
conductive metallic oxides with good thermal 
properties.\cite{Lee_Lowndes_et_al:2004}
In thin films, SRO is intensely investigated as a possible
route to the realization of novel field-effect devices.\cite{Ahn_et_al:2006,
Takahasi_et_al:2006}
In addition, it is of particular interest to the 
spintronic\cite{Zutic/Fabian/Sarma:2005,Awschalom_Flatte:2007} and 
multiferroic\cite{Spaldin/Fiebig:2005,Ramesh/Spaldin:2007} communities,  which 
have been recently energized by the possible device applications 
available from engineering interface phenomena.\cite{Ohtomo_et_al:2002,Ohtomo/Hwang:2004,Hwang_science:2006,Huijben_et_al:2006,Yamada_et_al:2004,
de_Teresa_et_al:1999,Brinkman_et_al:2007}
However, one limitation in the design of thin film oxide devices is the 
observation of increased resistivity in metal oxides as the film thickness 
decreases.
Such behavior is clearly present in ultra-thin films of SrRuO$_3$, where a 
metal-to-insulator (MI) transition\cite{Toyota_et_al:2005} occurs at 
four monolayers.
This substantial change in the electrical conductivity presents a 
serious challenge for device miniaturization.
In this work we explore the underlying physics of the thin film MI-transition, 
which to date remain to be understood.
The 3$d$ transition metal oxides (TMOs) are known to possess strong 
electron-electron correlation effects that can drive a system 
that should be metallic within a simple band picture into an insulating 
state.
Due to the large spatial extent of the 4$d$-orbitals in the ruthenates,  
correlation effects are anticipated 
to be less important as stronger hybridization provides more effective 
screening and a reduced Hubbard $U$ (Coulomb repulsion energy).
Many experimental studies have already addressed the degree of 
electron-electron correlation in SrRuO$_3$ including X-ray and ultraviolet 
photoemission spectroscopy,\cite{Kim_et_al:2004,Park_et_al:2004,
Toyota_et_al:2005,Siemons_et_al:2007} specific heat 
measurements,\cite{Allen_et_al:1996} infrared and optical conductivity
measurements,\cite{Kostic_et_al:1998} and transport experiments.\cite{Cao_et_al:1997}
For example, Kim and coworkers\cite{Kim_et_al:2004} use X-ray photoemission 
spectroscopy (XPS) to identify how such correlations 
change within the ruthenate family, and 
Toyota {\it et al.}\cite{Toyota_et_al:2005} use photemission spectroscopy (PES) 
to detail the metal-insulator transition in SrRuO$_3$ as a function of film 
thickness concomitant with the onset of magnetism.
In all of these studies, the general consensus is that electron correlation 
effects {\it do} play a role in determining the electronic structure 
of this itinerant ferromagnet, but to what degree remains unclear.
Furthermore, some theoretical investigations have begun examining
covalency,\cite{Maiti:2006} correlation\cite{Maiti/Singh:2005} and 
orbital ordering\cite{Jeng:2006} effects in bulk SrRuO$_3$.
The magnetic properties of SRO under epitaxial strain have also been 
investigated with first-principles techniques.\cite{Zayak/Rabe:2006}
In this work, first-principles density functional theory (DFT) calculations 
are performed, first to identify the degree of correlation in bulk SRO, and 
second to investigate the driving force for the metal-insulator transition 
in ultra-thin films.
We use two approaches to introduce correlation effects into 
the conventional band theory (local spin density) approach for treating the Ru 
4$d$-orbitals and their hybridization with O 2$p$-orbitals: the 
local spin density + Hubbard $U$ (LSDA+$U$), and the 
pseudopotential self-interaction corrected 
(pseudo-SIC) local spin density methods.
In addition we investigate two structural variants -- the ideal cubic perovskite
structure and the experimentally observed orthorhombic structure, which includes
tiltings and rotations of the RuO$_6$ octahedra.
By comparing the results that we obtained for both methods and both structure 
types, we are able to comment on the nature of the metal-insulator 
transition in ultra-thin films. 

\section{\label{sec:structure} Crystal Structure \& Magnetism}
\begin{figure}
\includegraphics[width=0.38\textwidth]{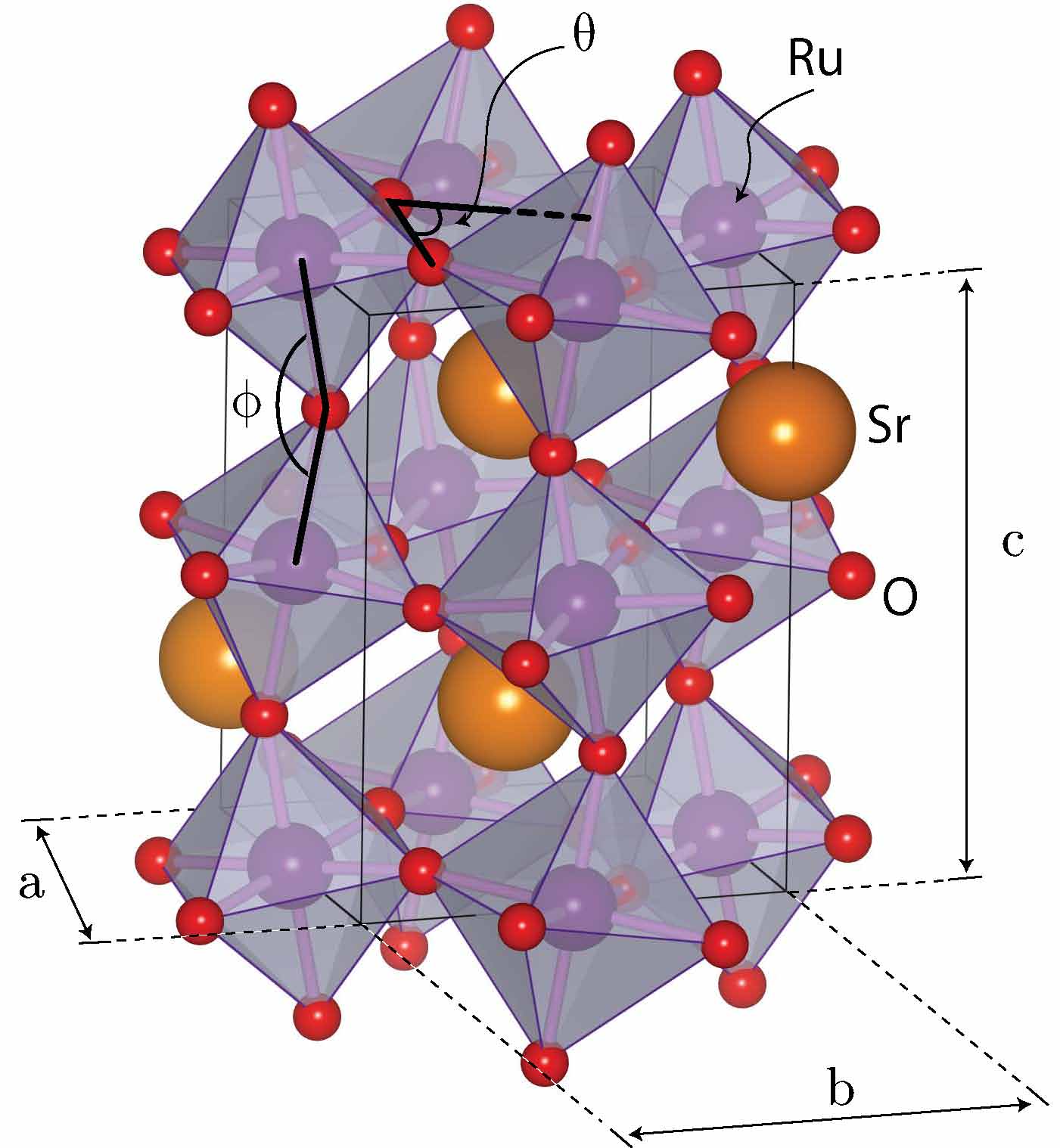}
\caption{\label{fig:structures} (Color online) The orthorhombic 
($Pbnm$) crystal structure of SrRuO$_3$.
The unit cell contains four formula units (f.u.) of the ideal cubic perovskite 
($Pm\bar{3}m$).
The structure is stabilized by shortening the Sr--O distance followed by 
a cooperative distortation of the RuO$_6$ octahedra to reduce the coordination 
volume of the Sr ions; this results in a smaller Ru--O--Ru bond angle, 
which in turn decreases the Ru 4$d$ bandwidth and metallicity.
The deviation in the structure from the high symmetry cubic state 
can be quantified using the tilting angle $(180^\circ-\phi)/2$ and 
the rotation angle $(90^\circ-\theta)/2$ of the oxygen octahedra.}
\end{figure}

The perovskite class of materials is described by a network of corner-sharing
$B$O$_6$ octahedra, in which the $A$-site cation is located at the center of 
a cube defined by eight $B$O$_6$ units.
The ideal perovskite is cubic (space group $Pm\bar{3}m$), however several 
modifications exist owing to the range of cation sizes that can be 
accommodated in the structure.
Deviations from the ideal cubic structure are defined by the Goldschmidt 
tolerance factor
\[
t^\prime = \frac{R_A+R_O}{\sqrt{2}(R_B+R_O)} \quad ,
\]
where $R_i$ is the radius of atom $i$, and can be attributed to the 
requirement to optimize the anion coordination 
about the $A$-site cation.\cite{Goldschmidt}
Using the Shannon-Prewitt radii for this compound, a predicted tolerance factor 
of $t^\prime$=0.908 is found, which is far from the ideal case $t^\prime$=1, 
suggesting that distortions should occur.
Indeed, SRO undergoes a series of structural transformations with temperature, from high 
symmetry cubic ($Pm\bar{3}m$, stable above 950~K) to tetragonal ($I4/mcm$, stable 
between 820~K and 950~K) to distorted orthorhombic structure ($Pbnm$) at low temperatures. 
The orthorhombic distortion from the ideal cubic can be described by the 
tilting of the RuO$_6$ octahedra in alternate directions away from the $c$ 
axis, and the rotation of the octahedra around the $b$ axis; in both cases
adjacent octahedra distort in the opposite sense (Figure \ref{fig:structures}).
The degree of tilting and rotation (as defined in Fig.\ \ref{fig:structures})  
of the octahedra are useful in describing the distortions in the oxygen 
network from the perfect cubic case.
A rotation angle of 7.56$^\circ$ and a tilting angle of 10.47$^\circ$ 
(corresponding to a Ru-O-Ru angle of 159$^\circ$) are found 
for $Pbnm$ SrRuO$_3$.
The structural changes reduce the hybridization between the Ru 4$d$ states and O 
2$p$ states and lead to a narrowing of the bandwidths (see Section \ref{RD}) compared 
with the ideal cubic case. 
Consequently the degree of correlation, described by $U/W$ where $W$ is the valence
bandwidth, is expected to be enhanced. 
Below approximately 160~K, SrRuO$_3$ exhibits strong ferromagnetic 
behavior, and has a measured Rhodes-Wohlfarth ratio\cite{Fukunagai:1994} 
($\mu_{\rm eff} / \mu_{\rm sat}$) of 1.3 suggesting that its magnetism can be 
well described by a localized $d$-electron model similar to the elemental 
ferromagnetic metals.
Within this model and under an octahedral crystal field, the $4d$-manifold 
splits into a threefold degenerate $t_{2g}$ subband that is lower in 
energy than the twofold degenerate $e_g$ band. 
Neglecting covalency, we would expect a spin-only magnetic moment of 
2~$\mu_B$,  corresponding to a low-spin state for the Ru$^{4+}$ ions 
($d^4: t_{2g\uparrow}^3, t_{2g\downarrow}^1$).
Experimentally, however, the moment is measured to be closer to 1.1~$\mu_B$/f.u., 
although values ranging from 0.9~$\mu_B$/f.u. and 1.6~$\mu_B$/f.u. have also
been reported.\cite{Kanbayasi:1976} 
(The spread in values is attributed to the large magnetocrystalline 
anisotropy of the material, and the difficulty in making large single-domain 
samples.)
First-principles calculations also report a magnetic moment ranging 
from 0.9~$\mu_B$/f.u. to 2.0~$\mu_B$/f.u.\cite{Singh_SrRuO3:1996,Allen_et_al:1996,Mazin/Singh:1997,Santi_Jarlborg:1997}
The reduced calculated magnetic moment in the solid compared to that in 
the free ion limit is due in part to the large spatial extent of the Ru 
4$d$ orbitals, which results in a significant overlap (hybridization) with the 
oxygen 2$p$.
Furthermore, due to the metallic character of SRO, an overlap of the 
majority and minority Ru 4$d$ bands occurs at the Fermi level; as a result
partial occupation of the minority band also leads to a reduced magnetic 
moment.
In this work, we examine the LSDA and ``beyond-LSDA'' electronic and magnetic 
properties of both the $Pbnm$ and $Pm\bar{3}m$ crystal variants. 
Metallicity and magnetism are both related to the $d$-bandwidth, which in 
turn depends on both correlations and structural properties such as
tiltings and rotations of the oxygen octahdera. 
Our goal, therefore, is to identify the relative contributions of 
electron-electron correlation effects and structural distortions in 
driving the metal-insulator transition in SrRuO$_3$ thin films.

\section{Theoretical Methods}
\subsection{LSDA}
Our initial electronic band structure calculations were performed within 
the local spin density approximation\cite{Kohn/Sham:1965} (LSDA) using both the 
{\sc siesta}\cite{Soler_et_al:2002,Siesta2,Ordejon95} and 
{\sc vasp}\cite{Kresse/Furthmueller_PRB:1996,Kresse/Joubert:1999}
 density functional theory (DFT) packages.
In each method we used the Perdew-Zunger\cite{Perdew/Zunger:1981} parametization of 
the Ceperley-Alder data\cite{Ceperley_Alder}for the exchange and 
correlation (XC) functional.
%

The core and valence electrons were treated with the projector-augmented-wave (PAW) 
method\cite{Bloechl:1994} for all calculations performed with 
{\sc vasp}.\footnote{We used 10 valence electrons for Sr (4$s^2$4$p^6$5$s^2$), 
14 for Ru ($4$p$^6$4$d^7$5$s^1$), and 6 for each oxygen (2$s^2$2$p^4$).}
Furthermore a plane-wave energy cutoff of 500~eV was used and found to produce
excellent convergence.
The three-dimensional Brillioun zone was sampled with a 
$12 \times 12 \times 12$  $k$-point Monkhorst-Pack mesh\cite{Monkhorst_Pack} 
for the cubic bulk (5 atom unit cell)  and thin film structures and a 
$12 \times 12 \times 10$ $k$-point sampling mesh for the bulk orthorhombic 
structure (20 atom unit cell).
For the orthorhombic films we used a  $12 \times 12 \times 3$ $k$-point sampling. 
In all cases the tetrahedron method with Bl{\"o}chl 
corrections\cite{Bloechl/Jepsen/Andersen:1994} was used for the Brillouin zone 
integrations.
%

In the localized basis code {\sc siesta}, the core and valence electrons were 
treated with norm-conserving fully separable\cite{KleinmanBylander} 
Troullier-Martin\cite{troullier} pseudopotentials.\footnote{The electronic 
configurations for each atom is:
Sr 4$s^2$4$p^6$4$d^0$4$f^0$ (1.50, 1.50, 2.00, 2.00), 
Ru 4$s^2$4$p^6$4$d^7$4$f^0$ (1.30, 1.30, 1.40, 1.30), and
O  2$s^2$2$p^4$3$d^0$4$f^0$ (1.15, 1.15, 1.15, 1.50), where the cutoff radii 
for each orbital is given in parentheses.}
The localized atomic orbitals for each atom used a single-$\zeta$ basis set 
for the semicore states and a double-$\zeta$ for the valence 
states.
Total energies were computed on a uniform real space grid with a cutoff of 
800~Ry in order to reach comparable accuracy to the planewave code.\footnote{
The grid cutoff mentioned is not directly comparable to the 
plane-wave cutoff value used to represent the wavefunctions in a standard 
plane-wave implementation of DFT, it is rather used here to represent the 
density, and is typically four times larger than the wavefunction cutoff.}
The Brillioun zone of the cubic structure was sampled with a 
$26 \times 26 \times 26$ $k$-point Monkhorst-Pack mesh, while the $Pbnm$ structure was sampled
with a $15 \times 15 \times 12$ $k$-point mesh. 
Integrations were performed with a Gaussian broadening of 0.10~eV in all 
calculations.

The equilibrium lattice parameter for the cubic structure was found by fitting 
the total energy as a function of volume to the Murnaghan equation of state.
Excellent agreement was found between the two codes 
(Table~\ref{tab:Vol_data}), with a  slight underestimate of the experimental 
lattice constant typical for the LSDA.
\begin{table}
\begin{ruledtabular}
\begin{tabular}{ccc}
		&	\sc vasp	& 	\sc siesta	\\
\hline
$a/a_o$		&	0.98		&	0.98		\\
$B$~ (GPa)	&	200		&	219		\\
$B^\prime$	&	4.6		&	4.4		\\
Moment ($\mu_b$/f.u.)&	1.09		&	1.26		\\
\end{tabular}
\end{ruledtabular}
\caption{\label{tab:Vol_data} Results obtained for cubic SrRuO$_3$ within the 
local spin density approximation (LSDA) from the two codes used in this work. The 
equilibrium lattice constant relative to the experimental value 
($a_0=3.9735$~\AA) 
determined from high-temperature neutron diffraction 
data,\cite{Chakoumakos/et_al:1998} the bulk modulus $B$, the pressure 
derivative $B^\prime$, and the magnetic moment per formula unit are given for  
each code. Very good agreement is found between the two codes. The notation for the codes are as follows, 
{\sc siesta}: 
Spanish Initiative for Electronic Simulations with Thousands of Atoms local 
orbital code and {\sc vasp}: Vienna Ab-initio Simulation Package planewave 
code.
}
\end{table}
The cell parameters and atomic positions of the orthorhombic structure 
(Table \ref{tab:lattice_info}) were optimized by starting from the 
positions reported in Ref.\ \onlinecite{Zayak/Rabe:2006} and the ionic 
coordinates were relaxed until the Hellmann-Feynman forces on the atoms were 
less than 4~meV~\AA$^{-1}$.
\begin{table}
\begin{ruledtabular}
\begin{tabular}{llccc}%
Atom	&	Site	&	$x$	&	$y$	&	$z$	\\
\hline
Sr	&	$4c$	&	-0.0050 &	0.03039 &	0.25	\\
Ru	&	$4a$	&	0.5	&	0.0	&	0.0	\\
O(1)	&	$4c$	&	0.0650  &	0.4942  &	0.25	\\
O(2)	&	$8d$	&	0.7158	&	0.2834	&	0.0340	\\
\end{tabular}
\end{ruledtabular}
\caption{\label{tab:lattice_info}Calculated structural parameters for SrRuO$_3$ 
with the $Pbnm$ symmetry using {\sc vasp}. Our calculated orthorhombic lattice constants 
are $a=5.4924$, $b=5.4887$, and $c=7.7561$~\AA.} 
\end{table}
Investigations of the electronic structure of cubic SrRuO$_3$ have been described 
by several different 
groups\cite{Singh_SrRuO3:1996,Allen_et_al:1996,Zayak/Rabe:2006} within the 
LSDA; here we briefly summarize their conclusions and remark that our results 
are consistent with the earlier calculations.
A complete comparison of the electronic structure of cubic SrRuO$_3$ 
calculated with both {\sc vasp} and {\sc siesta} is made in Section \ref{RD}.
In all cases, a metallic ferromagnetic ground state is found to be stable, with 
strong Ru 4$d$ character at the Fermi level.
Substantial hybridization occurs between the O 2$p$ states and the Ru 4$d$-states 
and no energy gaps are observed in the densities of states.
The calculated magnetic moment is also always reduced from the fully ionic 
limit of 2~$\mu_B$.

\subsection{LSDA+$U$}
The first ``beyond-LSDA'' method that we use to treat the exchange and 
correlation (XC) within DFT is the  local spin density approximation with 
Hubbard $U$ (LSDA+$U$).\cite{Anisimov/Aryasetiawan/Liechtenstein:1997} 
Here we use the spherically averaged form of the rotationally invariant 
LSDA+$U$ introduced by Dudarev {\it et al.},\cite{Dudarev_et_al:1998} in 
which only one effective Hubbard parameter, $U_{\rm eff} = U - J$,
is used, where $U$ and $J$ are the spherically averaged Hubbard repulsion 
and intra-atomic exchange for electrons with the angular momentum of interest; 
in this case Ru 4$d$ states. 
We treat the double-counting term within the fully-localized-limit 
and note that this should more correctly describe the insulating
side of the metal-insulator transition that we study here; 
an improved description of the metallic side might be achieved using 
the recently introduced interpolation between the around-mean-field and 
fully-localized-limit extremes.\cite{Petukov_et_al:2003}
Within these approximations, the LSDA+$U$ correction to the LSDA potential
is
\begin{equation}
\Delta V(mm' \sigma) = - (U-J) \left( \rho^{\sigma}_{mm'} - \frac{1}{2}
\delta_{mm'} \right) \quad,
\label{FLL}
\end{equation}
where $m$, $m'$ are the orbital indices, $\sigma$ is the spin index, and
$\rho^{\sigma}_{mm'}$ is the orbital occupation matrix. 
The effect of the LSDA+$U$ correction given by Eq.~\ref{FLL} is particularly 
transparent in the limit of diagonal $\rho^{\sigma}_{mm'}$ with orbital 
occupancies 1 or 0: 
Occupied orbitals experience a potential which is lower in energy by
$(U-J)/2$ compared with the LSDA, and the potential for unoccupied orbitals
is raised by the same amount. 
In this study we varied $U_{\rm eff}$ from 0 to 6 eV for the Ru $d$-states, 
(the standard LSDA corresponds to a $U_{\rm eff}=0$~eV).
Structural optimizations were also performed within LSDA+$U$ approximation, 
however negligible structural changes compared with the LSDA were observed.

\subsection{Self Interaction Corrections}
Our second approach for extending the treatment of the exchange and
correlation is to correct for the spurious self-Coulomb and 
self-exchange interactions which arise within the LSDA, using the 
pseudopotential self-interaction corrected 
(pseudo-SIC) method.\cite{Filippetti/Spaldin:2003,
Vogel/Kruger/Pollmann:1996,Pemmaraju/Sanvito:2007}
These self-interaction errors are small for materials with delocalized 
electronic states, but can be significant in systems with localized 
electrons where the interaction of an electron with itself is large.
Since the SIC in a periodic, extended system is not uniquely defined,
many different methods have been proposed to remove SI in DFT calculations  
for solids (for a review see Ref.~\onlinecite{Stengel_Spaldin:2008}).
The procedure followed in the pseudo-SIC method is to:
\begin{enumerate}
\item Project the occupied Bloch states onto the basis of the pseudo-atomic
orbitals.
\item Correct the potential for each Bloch state by the SIC for the 
pseudo-atomic orbital weighted by the projection, and scaled to account for the
relaxation energy.
\end{enumerate}
Note that only the valence bands are corrected since the empty 
conduction bands, derived from orbitals where the occupation numbers are 
close to zero, are not self-interacting. 
This is in contrast to the LSDA+$U$ method, in which the occupied bands 
are lowered in energy and the unoccupied bands raised.
In principle, however, the two formalisms would yield equivalent results
if a suitable $U-J$ (corresponding to the SIC energy) were applied to 
all orbitals in the LSDA+$U$ calculation. 
Indeed, whether the deficiencies of LSDA for strongly correlated systems 
derive from the absence of Hubbard $U$ or the self-interaction error or both 
remains an open question.
The pseudo-SIC method has some advantages over LSDA+$U$, since it does
not require a choice of which orbital to correct, nor of the $U$ or $J$
parameters, and it can be applied readily to both magnetic and non-magnetic 
systems. 
We note that this is the first application of pseudo-SIC to an
itinerant-correlated system, so the comparison with our LSDA$+U$
results provides a test for the pseudo-SIC method.

\section{\label{RD}Results \& Discussions}
\begin{figure}
\includegraphics[width=0.45\textwidth]{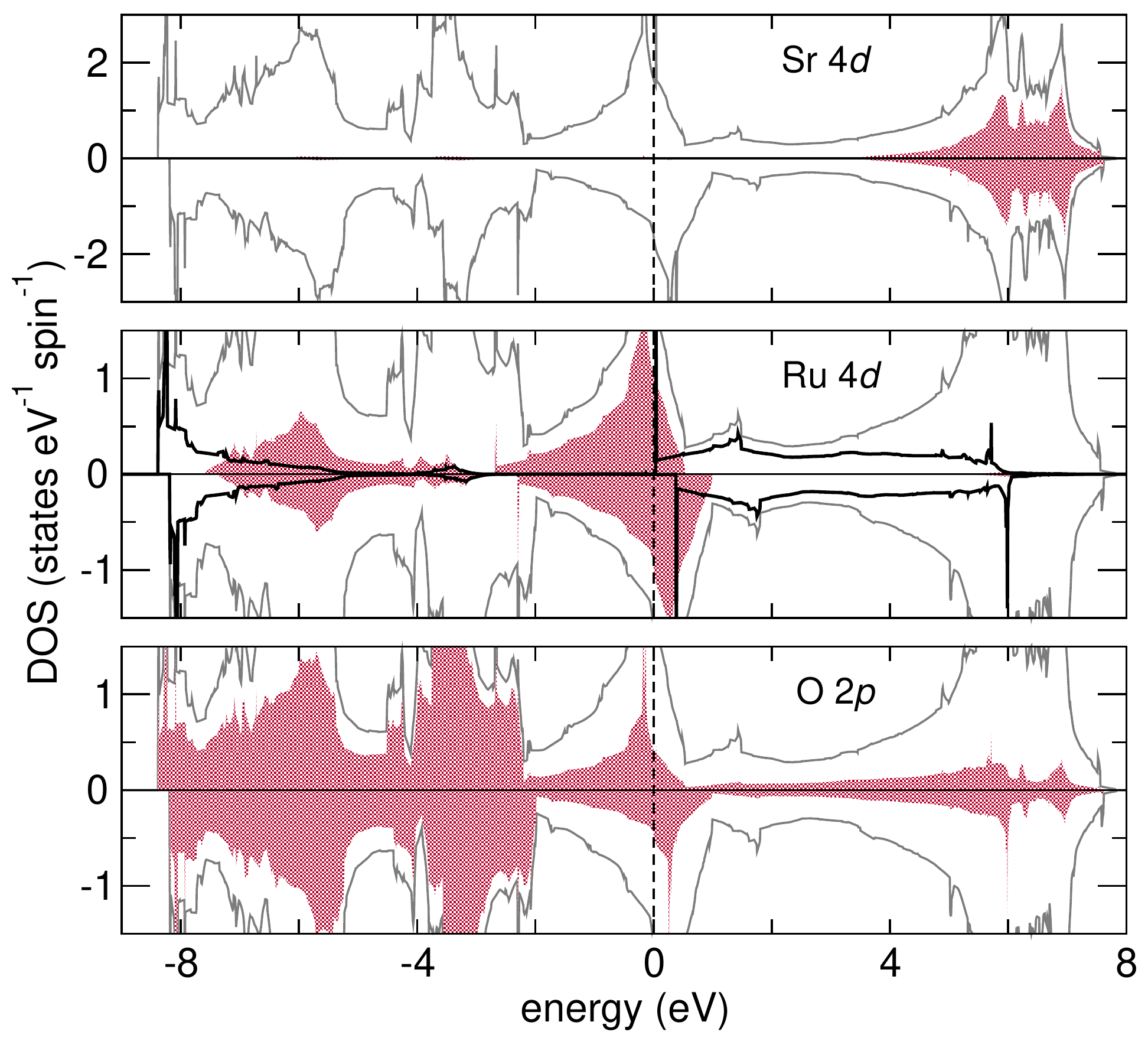}
\caption{\label{fig:cubic_dos}(Color online) The total (gray line) and 
partial (shaded) spin-resolved densities of states for 
cubic SrRuO$_3$ calculated within the LSDA using {\sc vasp}. ({\sc upper}) 
Sr 4$d$ states, ({\sc middle}) Ru 4$d$ states [the $t_{2g}$ 
and $e_{g}$ (unshaded, bold line) symmetries are shown], and 
({\sc lower}) O 2$p$ states.
The dashed line at 0~eV denotes the Fermi level.}
\end{figure}

As we have mentioned, the electronic structure of SrRuO$_3$ has been 
investigated previously using the 
LSDA.\cite{Singh_SrRuO3:1996,Allen_et_al:1996,Zayak/Rabe:2006}
Here we first revisit the LSDA with our own calculations, with 
an emphasis on understanding
discrepancies between the experimental measured photoemission 
results\cite{Kim_et_al:2004,Toyota_et_al:2005} and the calculated 
LSDA electronic structure of the orthorhombic material.
We then extend our study to the two ``beyond LSDA'' approaches to examine 
correlation effects in {\it  bulk} $Pbnm$ and $Pm\bar{3}m$  SrRuO$_3$.
Finally, we examine unsupported {\it films} of SrRuO$_3$ in both structures 
in order to analyse the nature of the experimentally
observed metal-insulator phase transformation.
{\it Cubic LSDA$\quad$} 
The total energies were calculated for both the ferromagnetically ordered
and non-magnetic states of cubic SrRuO$_3$ using the optimized lattice parameters.
With both electronic structure codes the ferromagnetic groundstate is always 
found to be lower in energy ({\sc vasp}: 11.5~meV, {\sc siesta}: 31.6~meV).
Both codes yield very similar electronic structures. The density of states
obtained using the {\sc vasp} code is shown in Figure \ref{fig:cubic_dos}.
The valence band is composed largely of O 2$p$ states hybridized
with Ru 4$d$ states, with oxygen states predominately found in lower regions of
the valence band and Ru states dominating at the Fermi energy. 
The large peak in the DOS near the Fermi level is caused by the fairly 
flat Ru $t_{2g}$ bands near the Fermi level while the strongly 
hybridized $e_g$ orbitals form broader bands at the bottom of the valence 
and conduction bands.
The Sr $4d$ states are found around 5~eV above the Fermi energy.
The exchange splitting causes an energy shift between the majority 
spin and minority spin states; at the $\Gamma$-point a splitting of 
$\approx$0.50~eV is observed in the Ru 4$d$ states, and of 
$\approx$0.20~eV in the O $2p$ states.
The calculated magnetic moments per formula unit are found to be 1.09~$\mu_B$ 
({\sc vasp}) and 1.26~$\mu_B$ ({\sc siesta}).
With both implementations of DFT, approximately 70\% of the moment is 
found on the Ru atoms, with the remaining distributed about the 
oxygen network.
The slight enhancement of the magnetic moment calculated with {\sc siesta} 
compared to {\sc vasp} is due to {\sc siesta}'s slight downward shift in 
energy of the Ru $t_{2g}$ band relative to the Fermi energy.
We note that the consistency between the two DFT flavors is essential to our 
later discussion of the effect of electron-electron 
correlations in the electronic structure of SrRuO$_3$, since the LSDA+$U$ 
approach has been implemented in the {\sc vasp} code, and the pseudo-SIC 
method in the {\sc siesta} code.
{\it Orthorhombic LSDA$\quad$} 
Using the optimized LSDA lattice parameters for the $Pbnm$ structure, we 
find that the ferromagnetic ground state is 6.34~meV/f.u.\ lower in 
energy than the constrained paramagnetic 
structure, and additionally is 188~meV/f.u.\ ({\sc vasp}) and 
150~meV/f.u.\ ({\sc siesta}) lower in energy than the 
ferromagnetic cubic phase.
This energy stabilization can be associated with the oxygen octahedral 
tiltings and rotations, and agrees well with previous first-principles 
studies\cite{Singh_SrRuO3:1996,Zayak/Rabe:2006} that used experimental 
lattice parameters\cite{Jones:1989} (the LSDA underestimates the lattice
parameters by only about 1\%).
Also it is consistent with the experimental observation of 
ferromagnetic SrRuO$_3$ in the distorted GdFeO$_3$ structure.\cite{Jones:1989}
\begin{figure}
\includegraphics[width=0.45\textwidth]{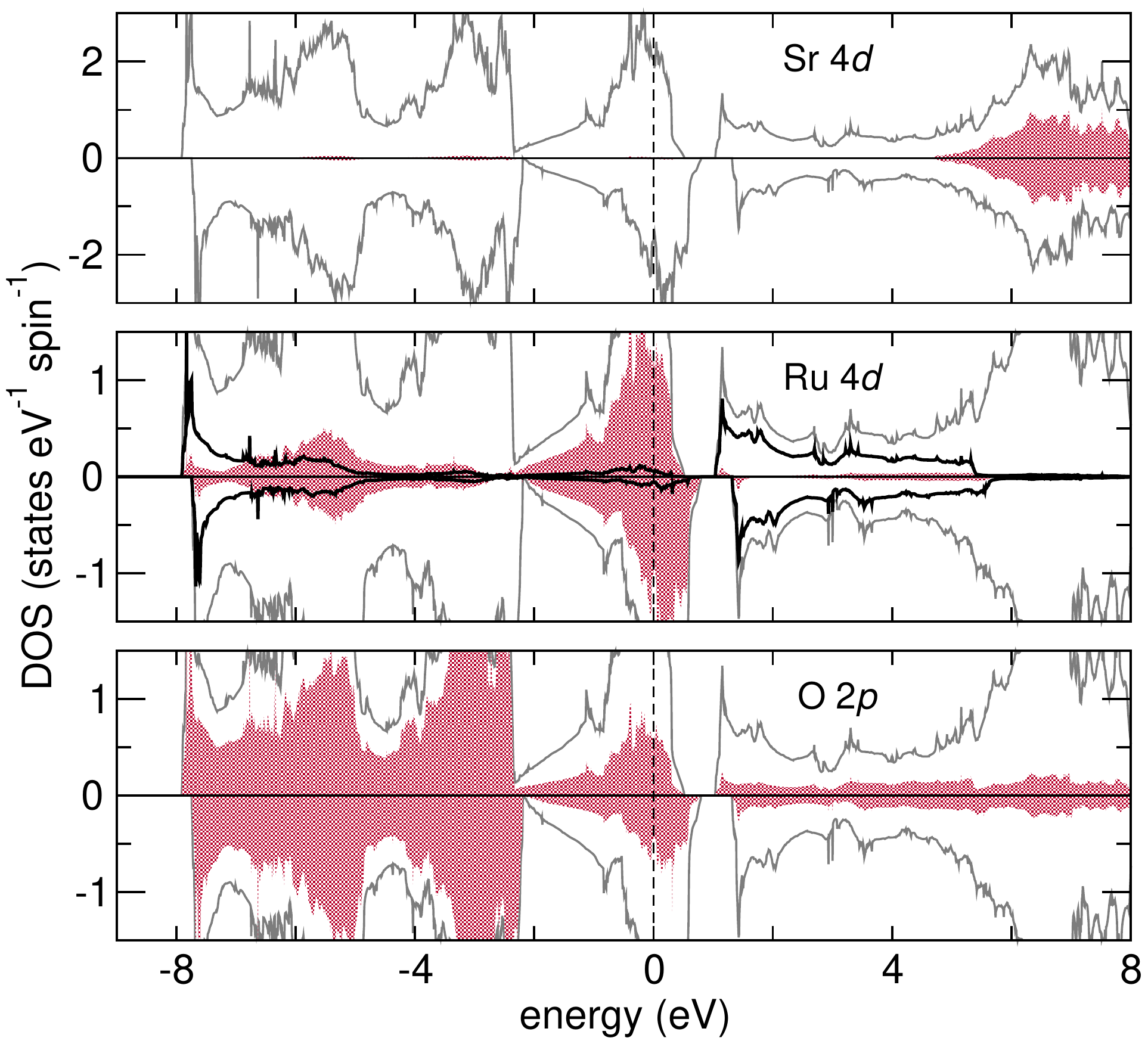}
\caption{\label{fig:ortho_dos}(Color online) The total (grey line) and partial 
spin-resolved (shaded) densities of states for orthorhombic SrRuO$_3$ calculated 
within the LSDA using {\sc vasp}. 
({\sc upper}) Sr 4$d$ states, ({\sc middle}) Ru 4$d$ states 
[the $t_{2g}$ and $e_{g}$ (unshaded, bold line) symmetries are shown], and 
({\sc lower}) O 2$p$ states.
}
\end{figure}
The (P)DOS for the orthorhombic structure are shown in Fig.\ \ref{fig:ortho_dos}, 
and can be seen to be similar to those of the cubic structure discussed earlier 
(Fig.~\ref{fig:cubic_dos}).\footnote{
Although in this symmetry the RuO$_6$ cages are rotated, we retain the 
standard spherical harmonics for the 4$d$ orbitals described in terms of 
a octahedral crystal field for a cubic perovskite.
The transformation from the cubic to orthorhombic $d$ orbital reference frame 
requires a rotation of $\pi/4$ about the [001]-direction; 
additionally, the octahedral units retain almost all of their 
integrity, i.e.\ the apical and axial Ru-O bond lengths are identical 
within 0.01~\AA.
}
Consistent with the reduction in Ru 4$d$ -- O 2$p$ overlap resulting from the
tiltings and rotations, the bandwidths are slightly narrower in the orthorhombic
structure, with the $t_{2g}$ bandwidth reduced by 0.35~eV, the $e_g$ by 1.5~eV 
and the O 2$p$ by 0.60~eV. 
This results in a pseudo-gap opening in the minority 
$e_g$ bands at $\approx$-2.2~eV and a 0.20~eV gap opening $\approx$0.80~eV above 
the Fermi level.  
Interestingly, the Ru $4d$ exchange splitting is reduced slightly to 0.30~eV
at $\Gamma$. 
This is accompanied by a reduction in the magnetic moment compared 
with the cubic structure, to 0.79~$\mu_B$/f.u.\ ({\sc vasp}) or 0.92~$\mu_B$/f.u.\ 
({\sc siesta}).
As noted, the spontaneous magnetization in the bulk (films) has been 
reported to be near 1.6~$\mu_B$ (1.4$\mu_B$), the LSDA underestimate
is likely the result of the usual LSDA overbinding leading to enhanced
Ru 4$d$ -- O 2$p$ covalency.
Note that this underestimate as compared to experiment\cite{Kanbayasi:1976} 
is significantly larger than that usually found for $3d$ 
transition metal oxides. 
It is further notable, because the orbital angular momentum is expected to be 
strongly quenched for 4$d$ orbitals due to the cubic crystal field. We comment on 
the effect of including spin-orbit coupling later.
Finally for the LSDA section, we compare our first-principles LSDA results 
with recent photoemission spectroscopy (PES) 
data\cite{Park_et_al:2004,Okamoto_et_al:1999,Fujioka_et_al:1997} 
with the goal of identifying which features are driven by correlation. 
%
In an ideal single-electron picture, the measured PES would consist of 
narrow peaks corresponding to the energies required to excite non-interacting
electrons from the valence band into the continuum.
However, the photoemission energies are more accurately interpreted as 
differences between two many-body $N$-electron states: the ground state, 
and the excited state with a photoelectron and hole.
The effect of the many-body interactions is to broaden the one-electron peaks 
and shift spectral weight into so-called quasiparticle peaks. 
The strongest reduction in spectral weight from correlation effects occurs 
from so-called coherent peaks near $\epsilon_F$, and is accompanied by transfer 
of the spectral weight to higher energy features (incoherent peaks).
Redistribution of the incoherent spectral weight into a well-defined 
satellite structure is indicative of strong correlations, whereas a 
redistribution into the background spectral distribution with a 
renormalization of the bandwidth indicates weak correlations.
%
%

\begin{figure}
\includegraphics[width=0.35\textwidth]{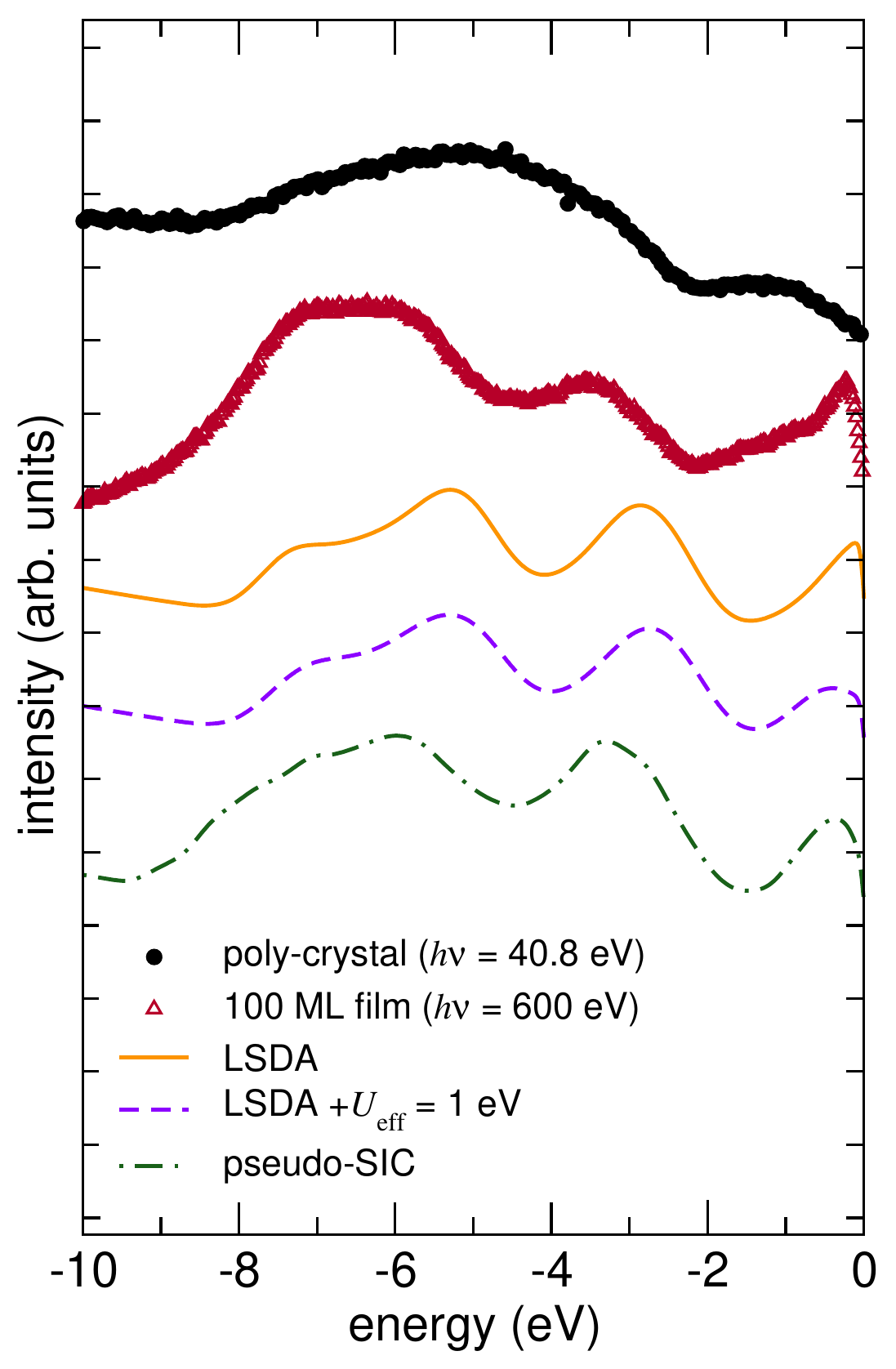}
\caption{\label{fig:ortho_u_exp}(Color online)
The experimental PES spectra for bulk polycrystalline\cite{Okamoto:private} 
SrRuO$_3$ (filled circles) and 100 monolayer SRO film\cite{Toyota:copyright} 
(triangles) grown on SrTiO$_3$ are compared to the calculated LSDA(+$U$) and 
pseudo-SIC densities of states. 
The calculated DOS are broadened with an energy dependent Lorentzian 
(FWHM = $0.1\left|\epsilon - \epsilon_F\right|$~eV) and a Gaussian 
function (0.34~eV FWHM). An energy dependent parabolic background has also 
been added.}
\end{figure}
In Figure \ref{fig:ortho_u_exp}, we show two experimentally measured spectra 
(see Refs.\ \onlinecite{Toyota_et_al:2005} and \onlinecite{Okamoto:private} 
for further information on the sample preparation) and our calculated LSDA
results. 
First we comment on the discrepancies between the bulk polycrystalline 
spectrum and that of the thin film.
Comparing the experimental spectra, we see that the thin film at 100 monolayers 
(which is representative of the bulk material) shows stronger coherent peaks
than the highly broadened structure of the polycrystalline sample.
In both cases however, the spectra are dominated by three principle features from
the Ru $t_{2g}$ states near the Fermi level and the O 2$p$ states between -8 and 
-2~eV.
If we look more closely around the Fermi level, the polycrystal sample has 
substantially reduced spectral weight, whereas near -1.3~eV it is enhanced 
compared to the film.
This shift in the spectral weight to the incoherent peak agrees well with 
previous experimental comparisons\cite{Kim_et_al:2005} made between SrRuO$_3$ 
films and polycrystals and is attributed to the creation of near surface 
states induced during {\it in situ} scraping and not due to intrinsic correlation 
effects.
Additionally the presence of grain boundaries and compositional defects are also 
known to yield reduced coherent peak features in polycrystalline samples.
For the remainder of this study, we therefore restrict our comparison of the PES data 
to the 100 monolayer  film, since it more accurately describes the intrinsic 
electronic structure.
By comparing to our calculated densities of states, we can assign the peak features 
in the PES data to the corresponding electronic states.
In order to make the comparison with our calculated DOS, we convolute an energy 
dependent Lorentzian function [full width at half maximum 
(FWHM) = $0.1\left|\epsilon - \epsilon_F\right|$~eV)] with the calculated DOS to 
account for lifetime broadening.
A Gaussian function with a FWHM = 0.34~eV is also used to account for the 
instrumental resolution, and an energy dependent parabolic background is also 
added.
We find that the bands between approximately -2.5~eV and -8~eV are due to the 
O 2$p$ states, with the peak of the non-bonding state centered at -3~eV. 
In the range between -8~eV up to the Fermi level are the occupied Ru 4$d$ states, 
with the $t_{2g}$ state lying across the Fermi energy beginning at -3~eV.
Some important discrepancies exist between our LSDA results and the spectroscopic data:
The Sr 4$d$ states are positioned approximately 1.5~eV lower in energy than 
is expected from the experimental spectra as determined in the BIS 
and XAS spectra (not shown).\cite{Park_et_al:2004,Okamoto_et_al:1999}
The spread of the O 2$p$ states is also underestimated by 2~eV. 
The most drastic difference, and one examined many times in the 
literature, occurs in the Ru 4$d$ states, where indications of correlations
are found. 
From Fig.\ \ref{fig:ortho_u_exp} it is clear that the  $t_{2g}$ states at 
$\epsilon_{F}$ are overemphasized in the LSDA calculation. 
Experimentally the Ru 4$d$ spectrum shows only a small weak coherent peak and 
is 1.5~eV broader than the LSDA predicts.
As mentioned, the signature of strong correlations as observed in the PES 
data is the strong renormalization (or even absence) of the quasiparticle 
peak near $\epsilon_F$ or large satellite peaks.
The large coherent $t_{2g}$ peak about 0.5~eV below the Fermi level has a 
substantial incoherent feature near 1.3~eV  and is good evidence for 
localized electronic states from strong correlation effects.
We also note that Santi and Jarlborg \cite{Santi_Jarlborg:1997} suggest 
that this suppression of the $t_{2g}$ states is possibly due to small 
matrix elements for the $d \rightarrow f$ and $d \rightarrow p$ transitions.
Using our two ``beyond-LSDA'' techniques to introduce correlation effects, 
we attempt to address whether this incoherent feature, not found in 
the LSDA calculations, is indeed due to strong electron-electron interactions.
More recent ultraviolet photoemission spectroscopy (UPS) 
data\cite{Siemons_et_al:2007} suggests that the Ru stoichiometry 
plays a significant role in determining the spectral weight and intensity of the 
$t_{2g}$ peak at the Fermi level.
For stoichiometric SrRuO$_3$ films, the spectral intensity near the Fermi level 
is in much better agreement with the calcuated PDOS, while Ru deficient samples 
have a reduced intensity.
These facts suggest that previous comparisions may have been made with 
non-stoichiometric Ru samples, which may be caused by a high partial pressure 
of oxygen during growth.
Although the Ru cation deficiency appears to explain the discrepancy with 
experimentally measured spectra, it is difficult to fully remove correlation 
effects that implicitly result from changes in stoichiometry, e.g.\ the 
$d$ bandwidth can be varied by changing the volume of the unit cell via 
changes in O--Ru--O bond angles which occurs upon Ru vacancy formation.
For example, Siemons and co-workers found the enhancement 
of an incoherent peak at approximately 1.5~eV below the Fermi level in ruthenium 
poor samples measured with ultraviolet photoemission spectroscopy.\cite{Siemons_et_al:2007}
Future first-principles calculations may be able to identify the role of these 
defects.
It is clear that  the LSDA poorly describes the electronic and magnetic 
structure of bulk SrRuO$_3$, and this suggests that some 
underlying physics is missing in the local spin density 
approximation.
We next explore two extensions of that description to include 
electron-electron correlation effects.

\subsection{``Beyond LSDA''}
\begin{figure}
\includegraphics[width=0.45\textwidth]{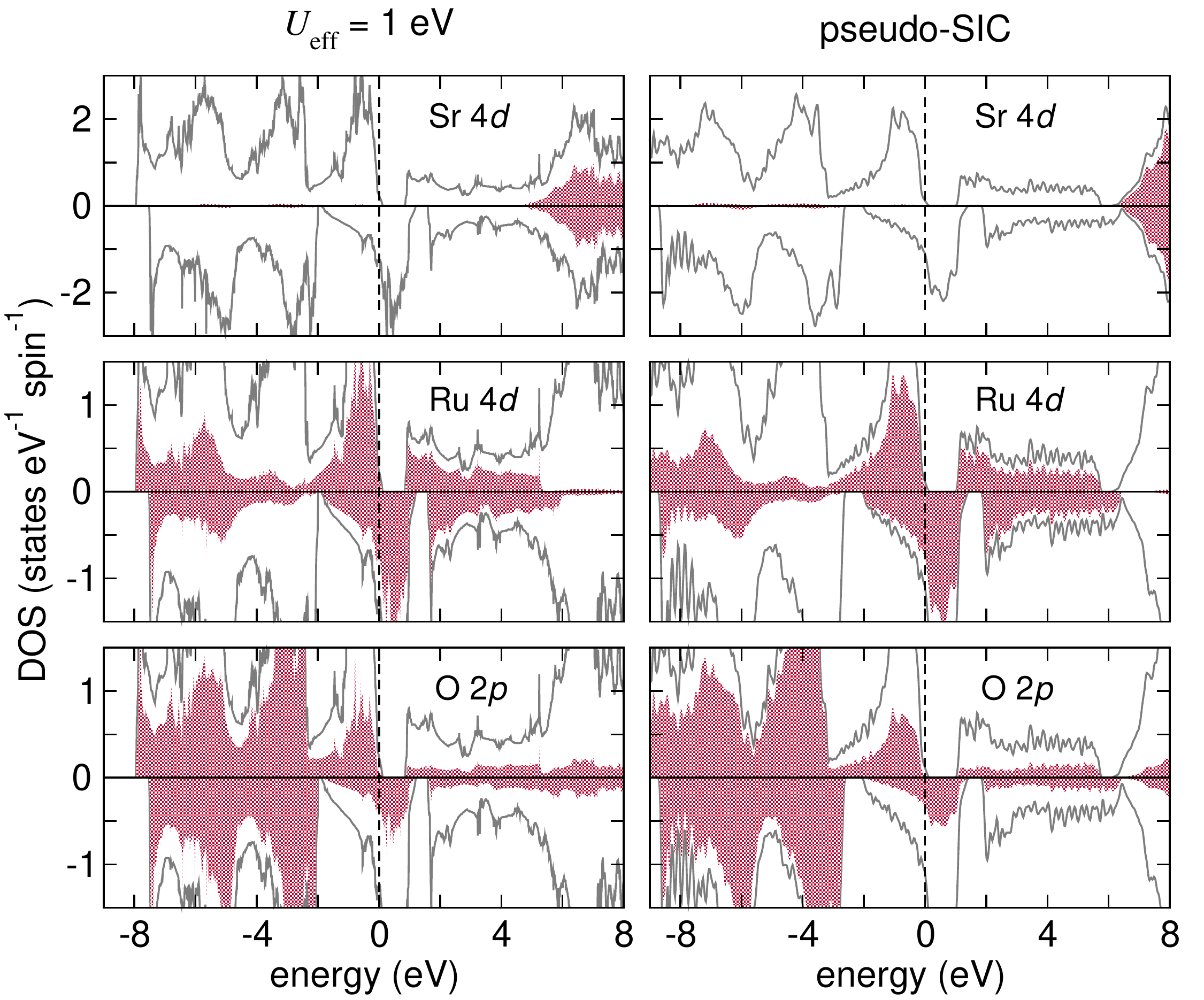}
\caption{\label{fig:asic_v_u1_pbnm}(Color online) 
The total (grey line) and partial 
spin-resolved (shaded) density of states for orthorhombic SrRuO$_3$ 
calculated with $U_{\rm eff}=1$~eV (left) and pseudo-SIC (right) are shown in 
each panel. ({\sc upper}) Sr 4$d$ states, ({\sc middle}) Ru 4$d$ states  
and ({\sc lower}) O 2$p$ states.} 
\end{figure}

We now explicitly add correlation effects into our electronic structure 
calculations for SrRuO$_3$ using the two methods outlined above.
The pseudo-SIC and LSDA+$U$ methods give very similar results when 
a $U_{\rm eff}=1.0$~eV is used. Therefore
we first compare the correlated and LSDA results,
and later point out the small differences between results from the two correlated
formalisms.
{\it Orthorhombic $\quad$} Figure \ref{fig:asic_v_u1_pbnm} shows the densities of 
states calculated with LSDA+$U$ with a $U_{\rm eff}=1.0$~eV and the pseudo-SIC 
method for $Pbnm$ SrRuO$_3$. Compared with the LSDA, the correlated bands are 
narrower, with energy gaps appearing in both spin channels.
The inclusion of correlations causes a 70\% drop in the total DOS at the 
Fermi level compared with LSDA, with the result that the contribution to the 
Ru 4$d$ states is almost entirely derived from the minority spin channel, 
and the material is close to half-metallicity.
This significantly enhances the magnetic properties compared to the LSDA,
increasing the magnetic moment per formula unit to around 1.0~$\mu_B$, and
enhancing the exchange splitting of the Ru $4d$ states at $\Gamma$ to 
0.45~eV ($\approx$0.30~eV in LSDA). 
In addition,
the peak positions of the correlation-included densities of states 
are in better agreement with the experimental spectra 
(Fig.~\ref{fig:ortho_u_exp}), although the intensity of the O 2$p$ peak at 
$\approx$ -7~eV is still too low compared with that of the Ru $t_{2g}$ peak.
The main difference between the electronic structures calculated with  
$U_{\rm eff}=1$~eV and the pseudo-SIC methods is a larger bandwidth of the
occupied orbitals, by $\approx$1.7~eV, in the pseudo-SIC calculation.
This has the greatest effect on the Ru 4$d$ bands and the oxygen 2$p$ 
bands near -8.0~eV.
As a result, the pseudo-SIC shows better agreement with the PES in the 
bandwidth for the O 2$p$ states between -8 and -4~eV, and the 
$t_{2g}$ states at the Fermi level have been more accurately surpressed.
\begin{figure}
\includegraphics[width=0.4\textwidth]{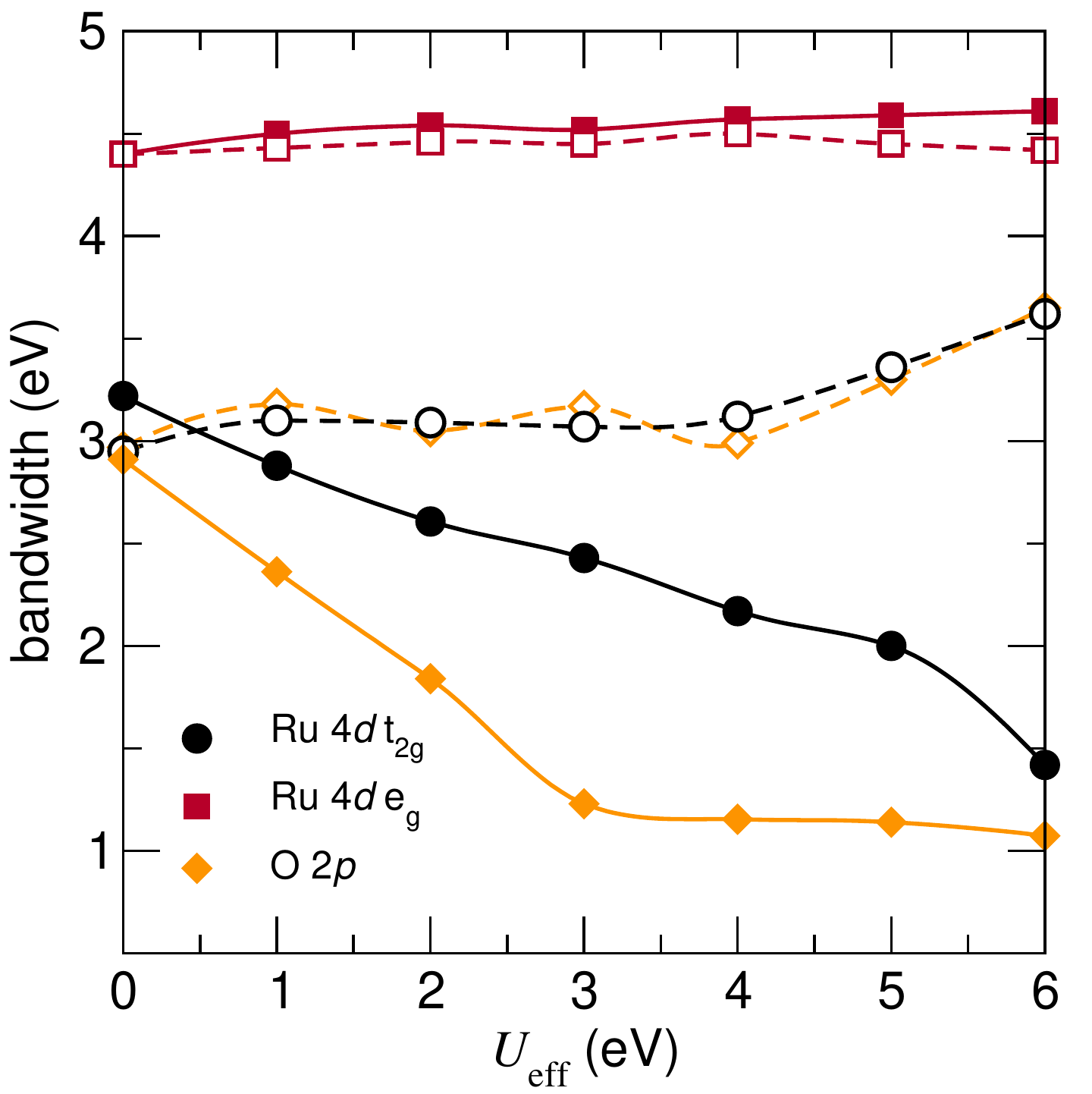}
\caption{\label{fig:ortho_U_W}(Color online) The orbital bandwidth dependence on 
 $U_{\rm eff}$ for orthorhombic SrRuO$_3$. The minority spin states are shown 
by the unshaded symbols and the lines are a guide to the eye. 
Using the pseudo-SIC method, the majority (minority) 
bandwidths are slightly larger than with the LSDA+$U$ method, but the 
relative ratios are consistent:
Ru 4$d$ $t_{2g}=3.30$ $(3.50)$~eV, Ru 4$d$ $e_g = 4.75$ $(4.60)$~eV, and 
O 2$p = 3.25$ $(3.50)$~eV.} 
\end{figure}

In order to understand how the hybridization changes as electron-electron 
correlation effects are included, we plot in Figure \ref{fig:ortho_U_W} the 
change in bandwidth for the Ru 4$d$ and O 2$p$ states as a function of 
$U_{\rm eff}$ for orthorhombic SrRuO$_3$.
As the amount of correlation is increased in the calculation through the 
$U_{\rm eff}$ term, the majority Ru $t_{2g}$ and O 2$p$ bandwidths are 
strongly reduced, and upon narrowing 
(both by approximately 1.80~eV) half-metallic behavior is observed for 
$U_{\rm eff}>2$~eV.
On the other hand, only a weak dependence in the orbital bandwidth is observed 
for the minority spin states.
The valence 
bandwidth never narrows sufficiently in the bulk material (due mostly to the 
insensitivity of the minority $t_{2g}$ bands to correlation effect) to open an 
insulating gap in both spin channels.
Since the half-metallic ground state that we find for $U_{\rm eff}>2$~eV 
is not observed experimentally, and motivated also by the observation of 
large magnetic anisotropy in Kerr rotation measurements\cite{Herranz_et_al:2005} 
on SrRuO$_3$ we have repeated our calculations with spin-orbit coupling (SOC) 
effects included.
\begin{figure}
\includegraphics[width=0.45\textwidth]{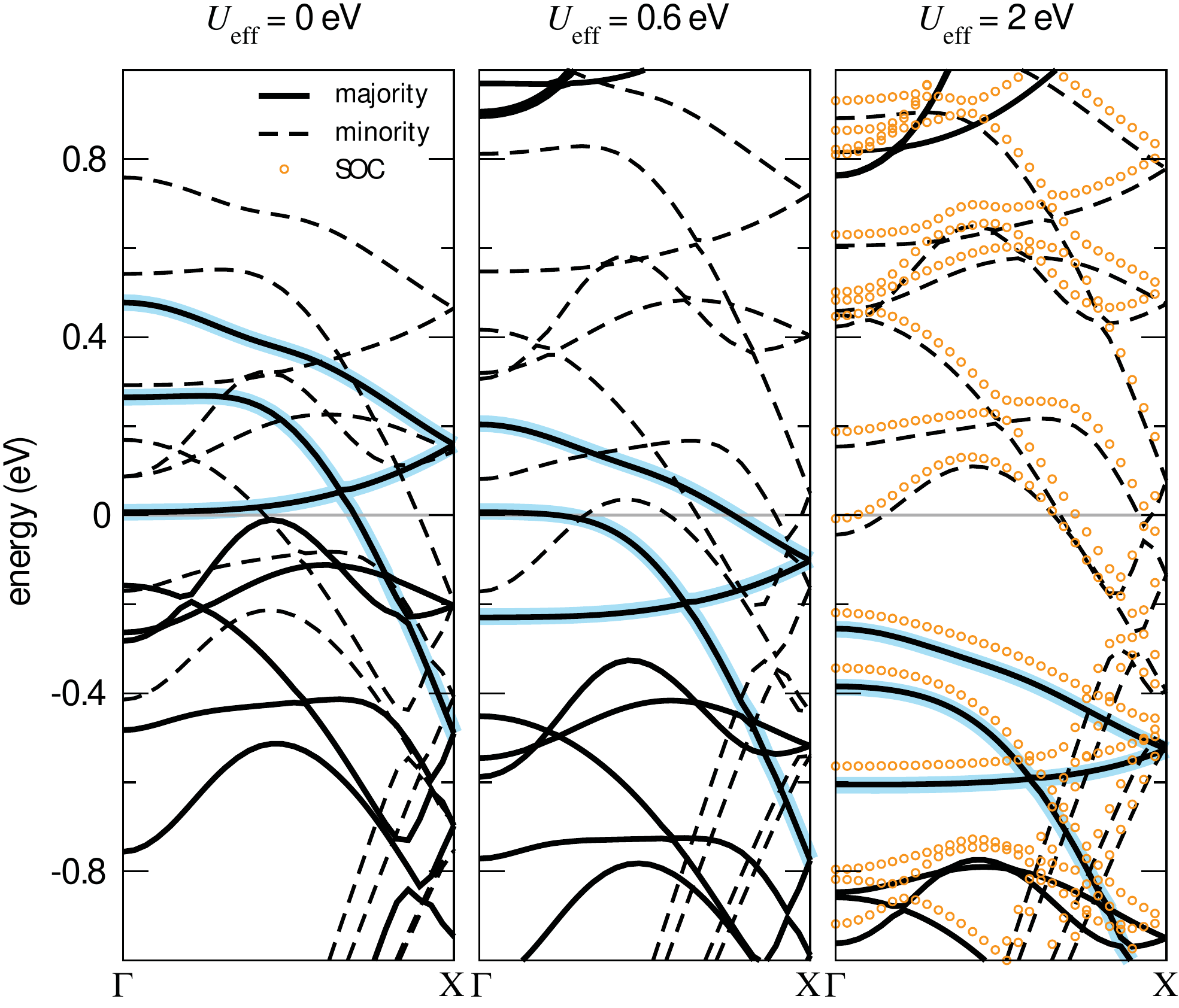}
\caption{\label{fig:ortho_bands}(Color online) 
LSDA+$U$ band structures along $\Gamma-X$ 
calculated with {\sc vasp} for ferromagnetic orthorhombic SrRuO$_3$.
The horizontal grey line marks the Fermi level.
Majority (minority) bands are shown as the bold (dashed) lines.
The highlighted bands indicate the filling of the majority $t_{2g}$ hole pocket, 
which gives rise to the observed half-metallicity at large Hubbard $U$ values.
For $U_{\rm eff}$ = 2 eV, the open circles show the results of calculations 
including spin-orbit coupling.}
\end{figure}
In Figure \ref{fig:ortho_bands} we plot the band structure along $\Gamma-X$ in 
the Brillouin zone as a function of increasing $U_{\rm eff}$.
We see that the $t_{2g}$ bands move down in energy with increasing 
$U_{\rm eff}$, forming a small hole pocket which becomes completely filled 
at $U_{\rm eff}$ = 2 eV giving the half-metallic behavior.
We note that without careful sampling of the Brillouin zone, this hole pocket 
is often missed, and half-metallic behavior can be prematurely predicted.
Furthermore, we superimpose the band structure calculated with 
$U_{\rm eff} = 2$ eV and spin-orbit coupling in Figure \ref{fig:ortho_bands}.
We find here that the degeneracy of the $t_{2g}$ 
bands is completely removed, and the highest occupied majority band is 
pushed only 0.05~eV higher in energy.
Furthermore, this splitting decreases at larger $U_{\rm eff}$ values.
These results indicate that the inclusion of spin-orbit coupling does 
not have a large effect on the band structure.
Although spin is strictly not a good quantum number, due to 
quenching of the angular momentum by the crystal field, the total angular 
momentum is well-approximated by the spin only component, and therefore the  
calculated proximity to half-metallicity in SrRuO$_3$ is robust to 
spin-orbit coupling effects.
Finally we investigate the enhancement of the magnetic properties
as $U_{\rm eff}$ is increased. 
\begin{figure}
\includegraphics[width=0.45\textwidth]{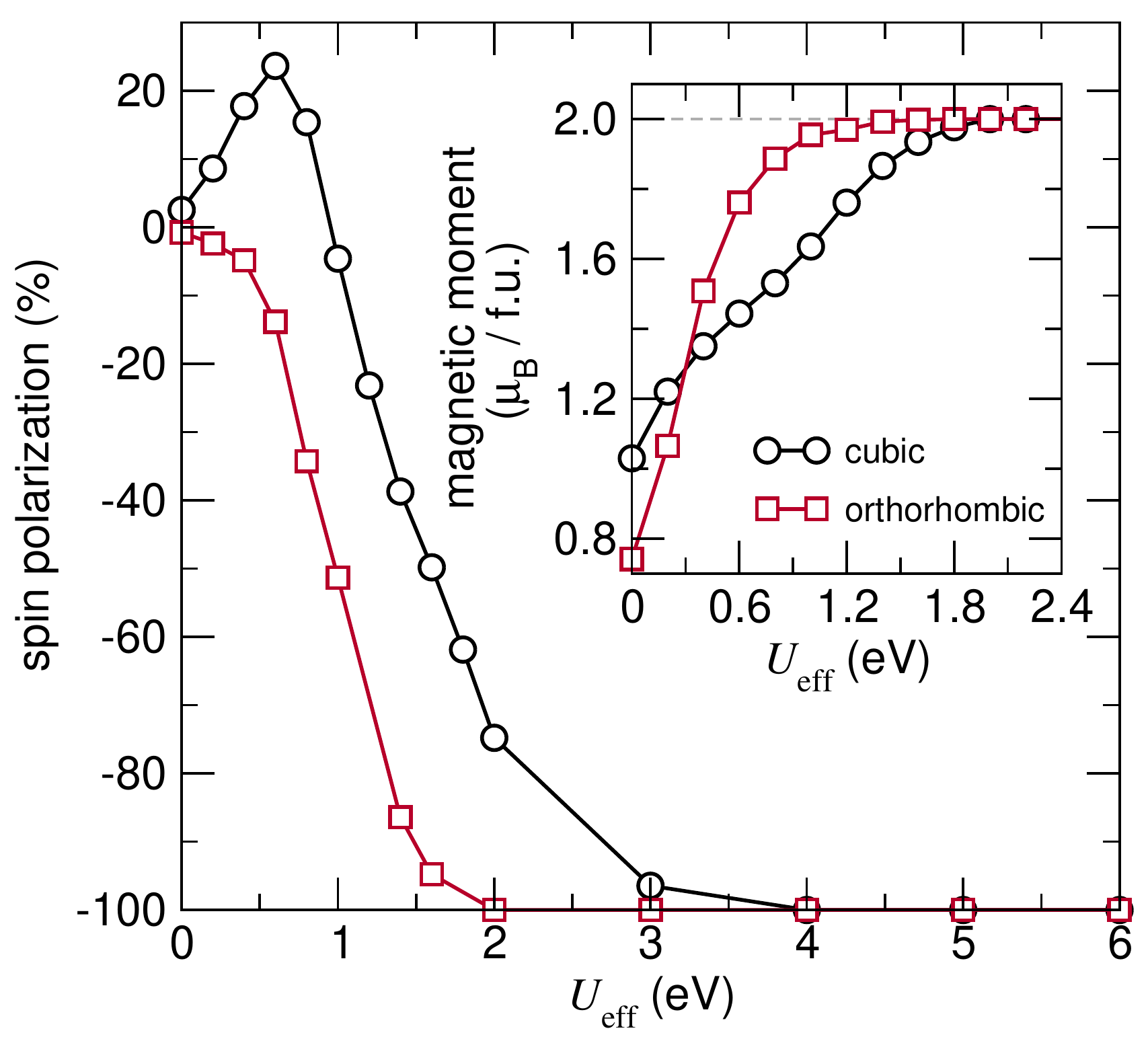}
\caption{\label{fig:mag_U}(Color online) 
Calculated spin polarization at the Fermi level 
($P_0^{\epsilon_F}$) for orthorhombic and cubic SrRuO$_3$ caculated with 
{\sc vasp} as a function of $U_{\rm eff}$.
The pseudo-SIC calculations give a spin polarization of +2.0\% and 8.8\% 
for each structure, respectively.
({\sc inset}) Calculated magnetic moment per formula unit for 
each structure type with LSDA+$U$. 
The pseudo-SIC calculations yield a magnetic moment of 1.99 and 
1.77$\mu_B$/f.u.\ for each structure, respectively.} 
\end{figure}
In Figure \ref{fig:mag_U} (inset) we show the magnetic moment per formula unit 
as a function of increased correlation for both crystal structures of SrRuO$_3$.
For example, the magnetic moment per Ru atom is found to be 1.97~$\mu_B$ 
with $U_{\rm eff}=1.0$~eV and 1.99~$\mu_B$ with the pseudo-SIC.
At $U_{\rm eff} \geq 2$~eV saturation of the moment occurs, and the 
localized ionic moment is observed (2.0~$\mu_B$).
{\it Cubic}$\quad$
We complete this discussion by describing the differences in the bulk 
cubic electronic structure with correlation effects added in order to 
isolate the contribution of the octahedral  
distortions in the orthorhombic structure to the bandwidth narrowing.
This analysis will provide the framework for exploring the MI-transition 
in the SrRuO$_3$ thin films.
Overall the weight and shape of the total density of states with correlations 
included is consistent with that calculated in the LSDA, with the exception 
that the large densities of states at the Fermi level 
(majority $t_{2g}$ states) are pushed lower in energy.
As was observed in the orthorhombic structure, a similar DOS is found between both 
correlation methods, and in general the occupied orbitals with the pseudo-SIC 
are broadened by 1.5~eV in energy compared to those calculated with the LSDA+$U$.
As in the orthorhombic structure the minority bandwidths in the cubic 
case are insensitive to the choice of $U_{\rm eff}$ while the majority O 2$p$ 
and Ru $t_{2g}$ bandwidths narrow considerably (1.2 and 1.8~eV, respectively).
Furthermore, the observed exchange splittings for the various states are overall 
larger for the cubic structure, and this is consistent with the electronic 
structures calculated with the LSDA.
The calculated magnetic moment for $U_{\rm eff}=1$~eV is 1.64~$\mu_B$ and 
agrees well with that from the pseudo-SIC method (1.77~$\mu_B$).
Again, for values of  $U_{\rm eff} > 2$ eV a half-metallic ground state 
becomes stable while with the pseudo-SIC a fully metallic ground state is 
always maintained.
To summarize the bulk SrRuO$_3$ results, each of the two ``beyond LSDA'' methods  
described here improve the description of the electronic and magnetic 
structure. However the precise experimental spectra are not fully reproduced 
although correct peak assignments can be made. 
The addition of a small Hubbard term $U_{\rm eff}=0.6$ (1.2)~eV for 
the orthorhombic (cubic) structure, or alternatively by correcting the SI 
error in LSDA, also improves the Ru $t_{2g}$ bandwidths with respect to 
experiment.\cite{Maiti/Singh:2005}
The total width of the O 2$p$ band structure is also increased to 
approximately 7~eV in agreement with the spectral weights.
We therefore suggest that SrRuO$_3$ can best be described as {\it weakly
strongly--correlated}.
Finally, as stated earlier, the intensity at $\epsilon_{F}$ has been 
decreased in comparison to LSDA, although it is still larger than 
experimently observered. 
\subsection{\label{sec:spin-polarization}Spin Polarization 
\& Transport Properties}
SrRuO$_3$ has been experimentally reported\cite{Worledge/Geballe:2000} to belong to 
the class of negatively spin-polarized materials-- characterized by a 
greater number of {\it minority} spins at the Fermi surface which are aligned 
anti-parallel to the bulk magnetization.
However, the magnitude of the spin polarization at the Fermi level remains 
controversial within the experimental community, due in part to the different definitions 
of the spin polarization (resulting from the different experimental techniques used 
to probe this quantity), as well as to difficulties in performing the experiments.
Furthermore, the theoretical community has also not converged on the
magnitude of the spin polarization, due to the sensitivity of the Ru $t_{2g}$ states near 
$\epsilon_F$ on the choice of exchange-correlation functional.
In this section, we perform first-principles transport calculations on 
orthorhombic and cubic SrRuO$_3$, and compare our results to the 
available data in the literature.
We also describe the various definitions of the spin polarization commonly used in 
the literature, and relate them to calculated {\it ab initio} quantities.
The spin polarization at the Fermi level $P_0^{\epsilon_F}$ can be 
calculated from the density of states at the Fermi level ($N_{\epsilon_F}$) by 
the following ratio, 
\begin{equation}
P_0^{\epsilon_F} = \frac{N_{\epsilon_F}^\uparrow-N_{\epsilon_F}^\downarrow}{
          N_{\epsilon_F}^\uparrow+N_{\epsilon_F}^\downarrow}\quad.
 \label{eqn:polarization}
\end{equation}
%
Using this definition with the LSDA, the sign of the spin 
polarization for orthorhombic SrRuO$_3$ is ambigous: We find that with the planewave 
code $P_0^{\epsilon_F}=-2.95\%$ while the local orbital code gives a positive spin 
polarization of 2.00\%.
In constrast, for the cubic structure we find a positive spin-polarization 
($P_0^{\epsilon_F}$) in both first-principles calculations, +1.3\% ({\sc vasp}) 
and +8.8\% ({\sc siesta}).
The reason for this discrepancy is the sensitivity of the exchange splitting 
of the Ru 4$d$ bands near the Fermi level to the structure.
The majority $t_{2g}$ band is positioned very close to the band edge, and 
its precise location is sensitive to the finer details of the DFT calculation. 
As a result, the large spread in the calculated spin polarization is not 
surprising, and since the magnitudes of the spin polarizations are small, 
changes of a few percent can give a change of sign.
When correlations are introduced, $P_0^{\epsilon_F}$ increases in magnitude 
signifigantly and is negative in all cases.
The spin polarization as a function of $U_{\rm eff}$ for the orthorhombic and 
cubic structures is plotted in Figure \ref{fig:mag_U}.
The spin-polarization calculated for the orthorhombic structure with the 
pseudo-SIC method is -85.7\%, compared with a value of +2.00\% when the SI 
error is not corrected.
This value agreesw with that obtained by the LSDA+$U$ when $U_{\rm eff}$=1.4~eV.
We previously found that smaller $U_{\rm eff}$ values optimize the
agreement between pseudo-SIC and LSDA+$U$ band structures and magnetic properties, 
suggesting that the pseudo-SIC transport results should be regarded as 
providing an upper bound on the magnitudes of spin-polarization.
For $U_{\rm eff}$ exceeding a critical value of 1.6~eV (3.0~eV) the half-metallic 
groundstate becomes the most stable solution for orthorhombic (cubic) structure, 
as an energy gap opens in the majority t$_{2g}$ band and $P_0^{\epsilon_F}$ 
reaches 100\%.
Despite being the most natural definition of spin polarization at the Fermi 
level, determining $P_0^{\epsilon_F}$ as defined in Eq.\ \ref{eqn:polarization}, 
is a non-trivial experimental process, since the spectroscopic measurements 
required typically have poor energy resolution.
As knowledge of the degree of spin polarization in a ferromagnet is crucial 
for its use in spintronics, several different experimental methods have been 
developed in order to determine this quantity.
The {\it transport} spin polarization can be defined as
\begin{equation}
 P = \frac{I^\uparrow - I^\downarrow}{I^\uparrow + I^\downarrow}\quad,
\label{eqn:current_polarization}
\end{equation}
where $I^\sigma$ is the spin dependent current. 
However $I^\sigma$ is not directly observable and must be determined indirectly. 
The transport spin polarization now depends on the experiment in question, and 
in particular whether the transport is in the ballistic or diffusive regime.
In the ballistic limit the current is proportional to $N_{\epsilon_{F}} \nu_{F}$, 
while for diffusive transport it is proportional to $N_{\epsilon_{F}} \nu_{F}^2$
(assuming both spin species have the same relaxation time), 
where $\nu_{F}^\sigma$ are the spin dependent Fermi velocities.
Therefore the transport spin polarization at the Fermi level can be redefined as
\begin{equation}
P_n^{\epsilon_F} = \frac{N_{\epsilon_{F}}^{\uparrow} \nu_{F}^{n \uparrow} - 
          N_{\epsilon_{F}}^{\downarrow} \nu_{F}^{n \downarrow}}{
         N_{\epsilon_{F}}^{\uparrow} \nu_{F}^{n \uparrow} + 
         N_{\epsilon_{F}}^{\downarrow} \nu_{F}^{n \downarrow}}\quad
\label{eqn:transport_pol}
\end{equation}
where $n=1$ for ballistic transport or $n=2$ for diffusive 
transport\cite{Mazin:1998,Coey:2004}.
If $n=0$, this definition reduces to that of the spectroscopic polarization, 
$P_0^{\epsilon_F}$.
An additional definition of polarization is used in Meservey-Tedrow style 
tunneling experiments.
Here the spin dependent DOS are weighted by their corresponding tunneling 
matrix elements.
Such an experiment has been performed for SRO and report approximately a 
-10\% spin polarization.\cite{Worledge/Geballe:2000} 
Inverse tunnel magnetoresistance measurements also agree that SRO is 
negatively spin polarized.\cite{Takahashi/Tokura_et_al:2005} 
This is in agreement with the majority of the calculations which find that SRO 
is a negatively spin polarized material at the Fermi surface.
The point-contact Andreev reflection (PCAR) technique, which is based 
on the process of Andreev reflection,\cite{Andreev:1964} and developed as 
an experimental method in the work of 
Soulen {\it et al.}\cite{Soulen_et_al:1998} and 
Upadhyay {\it et al.},\cite{Upadhyay_et_al:1998} has been used successfully 
to determine the magnitude of the transport spin polarization, although it 
is not sensitive to its sign. 
Experimental 
results\cite{Raychaudhuri/Beasley:2003,Nadgorny/Eom:2003,Sanders:2005} 
using this method report values ranging between 51\% and 60\%. 
It should be noted that in the Andreev experiment the polarization is not 
uniquely defined, in that it must be extracted from the data through a fitting 
procedure and involve terms that describe the transmittivity of the 
interface between the ferromagnet and the superconductor.
These parameters are typically difficult to determine precisely and consquently 
introduce further uncertainty in the experimental spin polarization.
Furthermore, it is important to note that in all PCAR experiments, it is 
necessary to establish whether the transport is in the ballistic, diffusive or 
intermediate regime (non-integer $n$) which ultimately depends on the 
transmittivity of the interface.
The experimental results for SRO are further complicated by the fact that the 
transport in the system has been measured in both regimes.
\begin{table}
\begin{ruledtabular}
\begin{tabular}{lcc}
	&	Orthorhombic	& Cubic  \\
\cline{2-3}
        & \multicolumn{2}{c}{$P_n^{\epsilon_F}$  \% (LSDA, pseudo-SIC)} \\
\hline
$n=0$	&	+2.00, -85.7	& +8.80, -16.1 \\
$n=1$	&	-1.44, -92.9	& -8.99, -50.9 \\ 
$n=2$	&	-15.1, -98.0	& -32.9, -79.5\\
\end{tabular}
\end{ruledtabular}
\caption{\label{tab:polarization} Transport spin polarizations calculated with 
{\sc smeagol}, according to the definition of Eq.\ \ref{eqn:transport_pol} 
using both the LDSA and pseudo-SIC.
Results for both the orthorhombic and cubic SrRuO$_3$ structures are included.}
\end{table}
\begin{figure}
\includegraphics[width=0.48\textwidth]{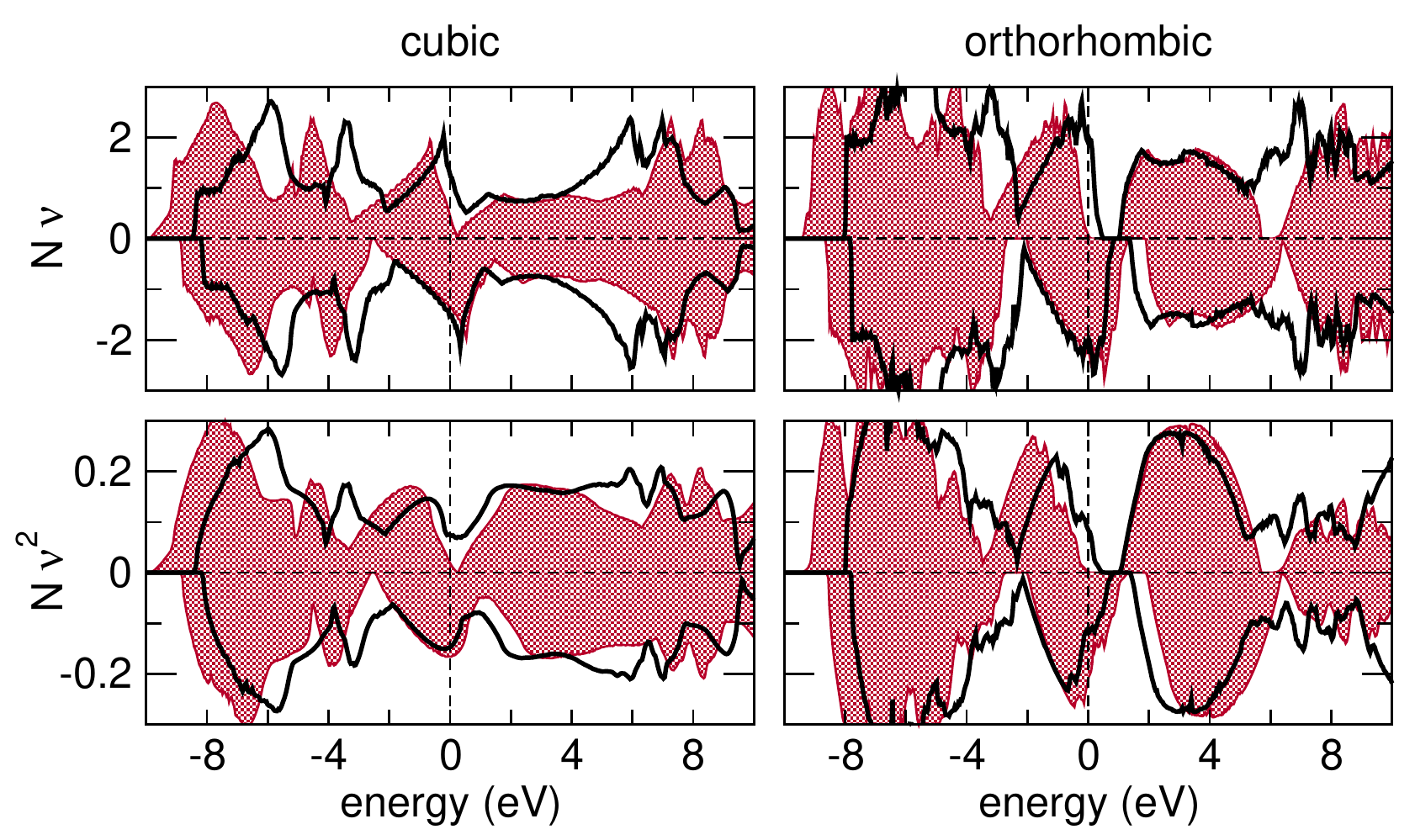}
\caption{\label{fig:transport}(Color online) Spin dependent transport 
coefficients, 
$N \nu$ and $N \nu^2$, calculated with both the LSDA (unshaded) and 
pseudo-SIC (shaded).} 
\end{figure}
To allow for a direct comparison with the PCAR experiments, the transport 
spin polarization in both the ballistic and diffusive limit was determined 
using the {\it ab initio} electronic transport code {\sc smeagol}.\cite{rocha:085414}
Here we calculated the transport at zero bias through both the orthorhombic 
and cubic structures,\footnote{
The Brillioun zone was sampled in these cases with a $20\times20\times1$ 
Monkhorst-Pack mesh for the orthorhombic structure and a $40\times40\times1$ 
mesh for the cubic structure.} and present the results in Table 
\ref{tab:polarization} and Figure \ref{fig:transport}.
The shortcomings of the LSDA in describing the spin polarization at the 
Fermi level in SrRuO$_3$ are again apparent. 
The highest spin polarization obtained with the LSDA for the orthorhombic 
structure is -15\% and it is obtained in the diffusive limit. 
This is notably smaller than the experimental PCAR results measuring the same 
quantity, $P_2^{\epsilon_F}$.
As shown in Fig.\ \ref{fig:transport}, on changing from $P_0^{\epsilon_F}$ to 
$P_2^{\epsilon_F}$ the polarization increases and becomes more negative.
Since the group velocity tends to zero at the band edge, and is often 
maximized at the band center,  higher powers of $n$ in $P_n^{\epsilon_F}$
suppress the contribution of the Ru 4$d$ states at the band 
edge while enhancing those at the band center.
From Figure \ref{fig:transport} it is then clear that the large 
negative polarization is a consequence of the center of the 
{\it majority} Ru 4$d$ band positioned approximately 1~eV below the Fermi 
level, while the {\it minority} Ru 4$d$ band center is aligned across the 
Fermi level.
Further enhancement is seen by introducing correlation; for example, by correcting
for the SI error, the spin polarization increases due to the reduction of the
number of majority Ru $t_{2g}$ states at the Fermi level. 
The correlated \emph{ab initio} calculations now give very high spin polarization, 
ranging between -85.7\% and -98.0\% where as the highest value achieved 
experimentally is just 60\%.
Qualitativity similiar results are found for the cubic structure, although the 
SIC in general has a smaller influence on the spin polarization. 
For example, $P_1^{\epsilon_F}$ goes from -8.99\% (LSDA) to -50.9\% (pseudo-SIC), 
while $P_2^{\epsilon_F}$ goes from -32.9\% to -79.5\%.

\begin{figure}
\includegraphics[width=0.48\textwidth]{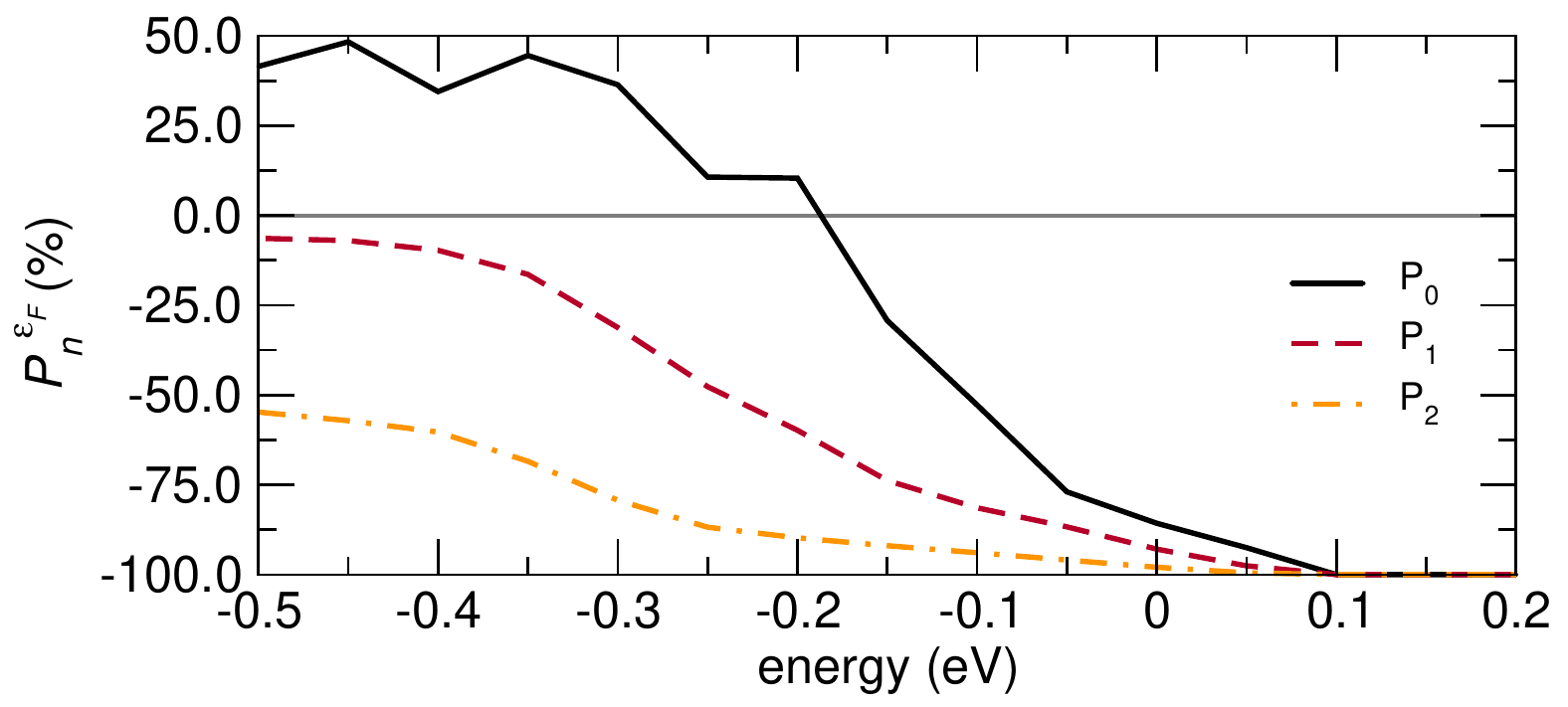}
\caption{\label{fig:polarization_ef}(Color online) Spin polarization defined 
according to Eq.\ \ref{eqn:transport_pol} as a function of distance 
from the Fermi energy (set to $0$~eV) and calculated with the 
pseudo-SIC for the orthorhombic structure.} 
\end{figure}
It is also useful to note the strong dependence of spin polarization on distance 
from the Fermi level. 
In Figure~\ref{fig:polarization_ef} we show for the orthorhombic structure 
that if the Fermi level is moved just 100~meV into the valence band, 
$P_1^{\epsilon_F}$ is decreased to -81.4\%, while moving $E_F$ by -200~meV 
decreases it further to -59.8\%, bringing it within the experimental range of 
values.
In practice this shift in the Fermi level can be realized by off-stoichiometric 
compounds such as those investigated by Siemons 
{\it et al.}\cite{Siemons_et_al:2007}
The discrepancy between the computational and experimental results could be then 
due to a number of factors: 
For example, there are several known limitations with PCAR including 
spin-flip scattering events which could drastically reduce the measured value of 
$P^{\epsilon_F}$, as well as the possible ambiguous fit of PCAR measurements 
to a multiparameter model.\cite{Taddei_et_al:2001}
We also note that spin-orbit coupling, which we did not account for in our transport
calculations, could reduce the spin polarization at the Fermi level. 
Despite these disparities, both \emph{ab initio} calculations and experiment show 
SrRuO$_3$ with a high negative spin polarization. 
As expected, LSDA underestimates the spin polarization at the Fermi level, 
whereas the inclusion of correlation through the correction of the SI error 
with the pseudo-SIC results in much better agreement between theory and experiment. 

\subsection{Thin films}
\begin{figure}
\includegraphics[width=0.48\textwidth]{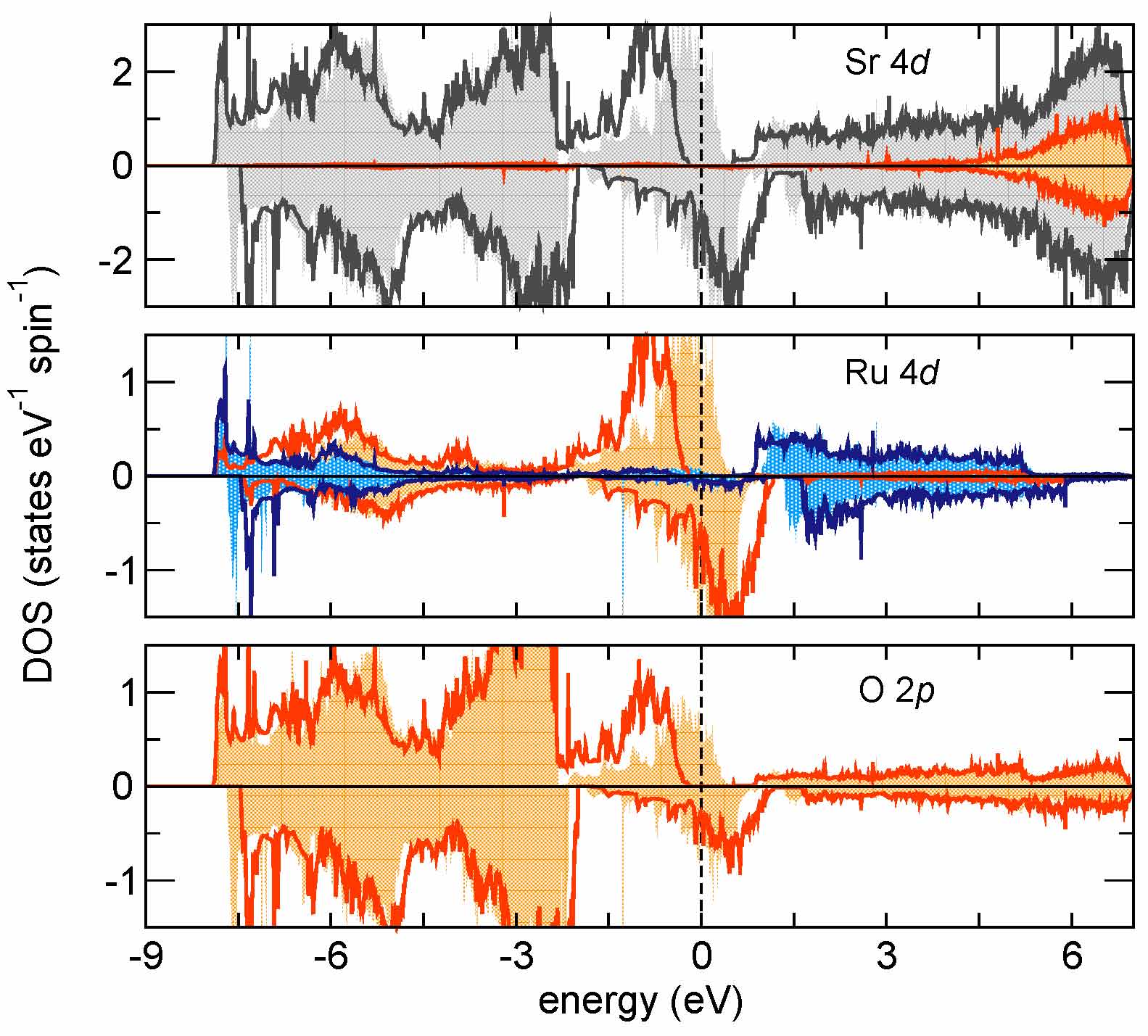}
\caption{\label{fig:3uc_slab_doses}The total and partial spin-resolved 
density of states for a three unit cell orthorhombic SrRuO$_3$ slab calculated 
with $U_{\rm eff}=0$~eV (shaded) and $U_{\rm eff}=2$~eV (unshaded, bold) 
are shown in each panel. ({\sc upper}) Total (grey) and Sr 4$d$-states, 
({\sc middle}) Ru 4$d$-states, $t_{2g}$ and $e_{g}$, and ({\sc lower}) 
O 2$p$-states.}
\end{figure}

The electronic and magnetic structure of epitaxially grown oxide multilayers 
can be tuned by controlling the film thickness.
In particular, it has been demonstrated that metallic SrRuO$_3$ can be 
transformed into an insulating state by growing films thinner than five
monolayers on SrTiO$_3$ substrates.\cite{Toyota_et_al:2005}
It was also found that the Curie temperature decreases with reduced film thickness, 
along with the disappearance of strong ferromagnetic order.
Photoemission experiments show a shift in the spectral weight to the incoherent 
peak features in the spectra, suggesting that these effects are a  
result of changes in electron-electron correlation effects.
With our first-principles techniques, we systematically investigate whether 
we can reproduce this transition purely from structural confinement, or by
also including correlation effects and/or the octahedral tiltings of the
orthorhombic structure. 
For the remainder of this section, we choose to include correlation with the 
LSDA+$U$ method, rather than the pseudo-SIC method, and 
note that from the discussion so far, both methods reproduce similar 
electronic structures.
{\it Cubic LSDA slabs$\quad$}
To investigate the effects of structural confinement on the 
metal-insulator transition, we first performed a series of slab 
calculations (from 1 to 5 unit cells thick) on cubic SrRuO$_3$ 
constrained to the calculated bulk equilibrium SrTiO$_3$ lattice parameter.
This is in part motivated by the fact that good epitaxy is made with the 
substrate surface, and that tilting of the octahedra may be suppressed.
Additionally, it is computationally more feasible to systematically investigate 
these smaller supercell slabs.
We discuss later the effect of including the octahedra tiltings in the orthorhombic 
thin films; we saw earlier that this structural effect is important in fully describing 
the subtle details of the electronic structure of SRO.
In all calculations the slabs were terminated with a SrO surface, to 
be consistent with that experimentally observed to be the most thermodynamically 
stable.\cite{Rijnders_CB_EOM_et_al:2004}
\begin{figure}
\includegraphics[width=0.45\textwidth]{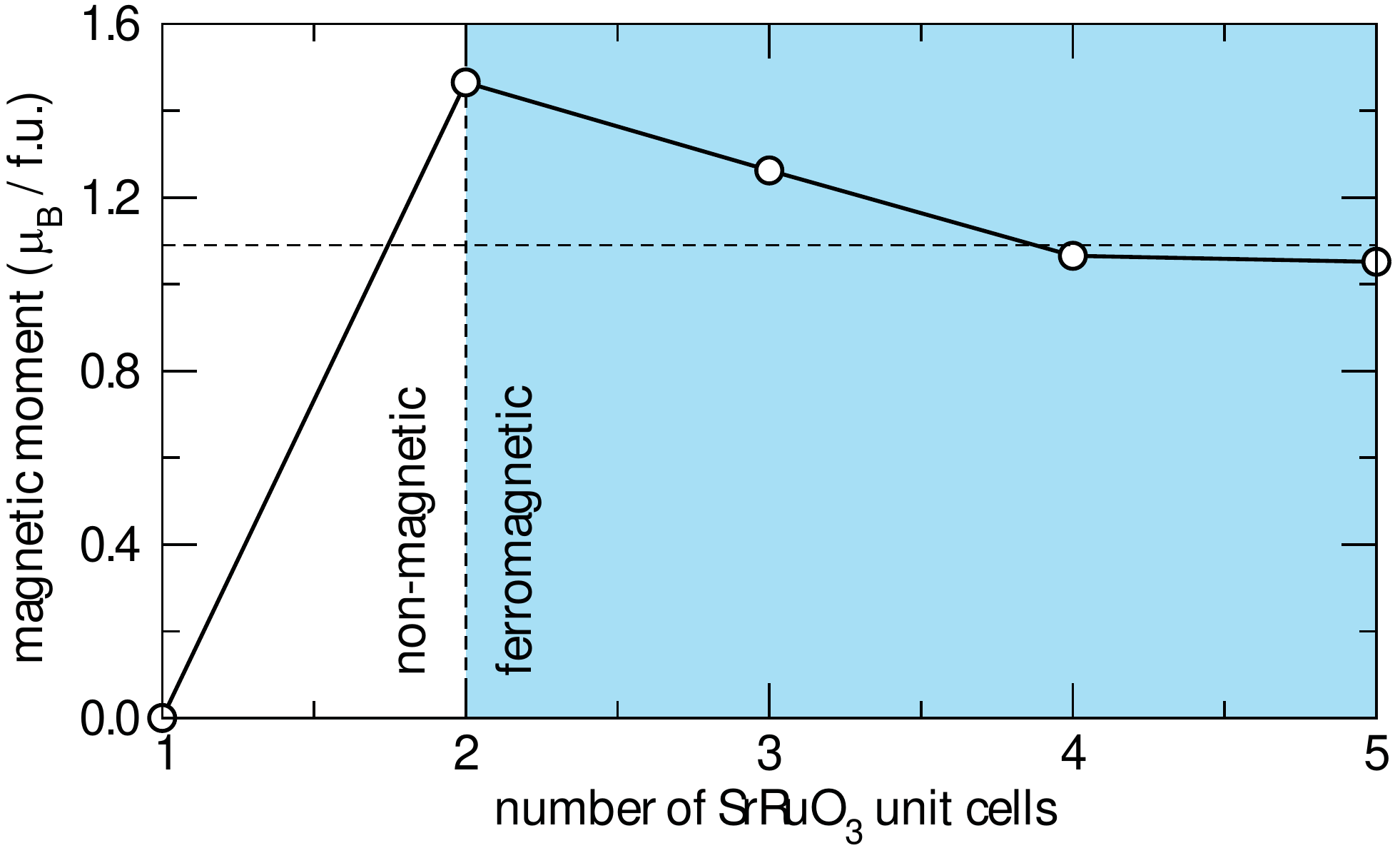}
\caption{\label{fig:slab_moments}(Color online) Magnetic moment dependence 
on slab thickness for cubic SrRuO$_3$. 
The bulk LSDA magnetic moment is shown as the dashed line.
The total energy differences are calculated for the spin-polarized 
films with respect to the non-magnetic ground state.}
\end{figure}
In Figure \ref{fig:slab_moments} we plot the Ru magnetic moment 
(per f.u.) as a function of increasing slab thickness.
We find that the LSDA films become non-magnetic below a critical thickness 
of only two monolayers; this is lower than the experimentally 
observed loss of the strong ferromagnetic order below six 
monolayers.\cite{Toyota_et_al:2006}
In addition, all of our calculations on the cubic films remain metallic down to 
one monolayer.
Experimentally the situation is different and insulating behavior is observed in 
heteroepitaxial thin films at six monolayers on SrTiO$_3$.
At one unit cell, where all atoms are surface-like, the magnetic moment is 
considerably suppressed from its bulk value, and the non-magnetic 
structure is actually lower in energy.
An enhancement in the magnetic moment is observed at two unit cells in 
thickness; the moment then decreases toward the bulk value as the film 
thickness grows.
For films larger than two unit cells, the ferromagnetic ground state is 
always found to be stable.
In a mean field theory approach, the energy difference between the ferromagnetic 
and the paramagnetic ground states $\Delta$ can be expected to be proportional 
to the Curie temperature $T_c$ according to $k_BT_c=\frac{2}{3}\Delta$.
For the four and five unit cell slabs, we find mean field $T_c$'s of 170 and 
120~K, respectively; these values are close to the experimental bulk
value of 160~K suggesting that even in these thin films the strong itinerancy 
remains. This is consistent with temperatue dependent magnetization data recorded 
on strained and free standing films\cite{Gan_et_al:1998} as well as on ultra-thin 
SrRuO$_3$ films.\cite{Toyota_et_al:2006}
Additionally,  the spin-polarization $P_0^{\epsilon_F}$ as a function of slab 
thickness (not shown) exhibits a large negative polarization at two unit cells, 
while a small positive spin-polarization is found with increasing thickness 
(consistent with our bulk spin polarization calculations).
Most importantly, the insulating state is not found in any of the cubic slab 
calculations nor is the non-magnetic ground state generally favored 
(the one unit cell case is an exception and is due to competing interactions 
from surface effects).
To better understand how the magnetism is distributed in the slabs, we have 
also calculated the layer-by-layer local density of states (LDOS). 
As in the bulk case, on average the majority of the spin ($\approx$65\%) is 
located on the Ru atom, with the remaining found on the oxygen network.
Interestingly, the Ru atoms closest to the surface layers experience a 
suppressed magnetic moment for each slab.
This is in contrast to most transition metal (non-oxide) ferromagnets, where often 
enhancement occurs due to a loss of coordination, weaker interatomic 
hybridization, and enhancement of the orbital angular momentum.
In this oxide, enhanced covalency at the surface layer may be responsible 
for the reduced magnetism.
Before adding correlation effects in the cubic slabs, we first discuss the 
changes in the electronic structure due to the thin film geometry.
The overall shape and weight of the density of states for the cubic slab 
and the bulk cubic LSDA calculation are very similar suggesting that confinement
effects are minimal.
The calculated exchange splittings are also similar with the exception that 
the Ru 4$d$ states are split by approximately 0.25~eV.
A small gap in the minority $e_g$ states opens at approximately {-4.30}~eV, 
and partial occupation of the majority $e_g$ states  occurs; these features 
are not observed in the bulk cubic LSDA calculation.
For a free standing, three unit cell film we find that the 
structure has a magnetic moment of 1.26~$\mu_B$ and a spin polarization of 
+13.2\% within the LSDA, both larger than the bulk cubic values of 
1.09~$\mu_B$ and +1.3\% respectively.
The increased positive spin polarization is a result of the band center of the 
minority Ru 4$d$ states shifting to higher energy in the thin films.
To summarize the results for the cubic SrRuO$_3$ thin films, we do not find a 
metal-insulator transition as a function of film thickness, although we 
do find a slightly enhanced magnetization.
We therefore are able to rule out the effect of dimensional confinement as the 
driving force for a metal-insulator transition.
{\it Cubic LSDA+$U$ slabs$\quad$}
We now examine the effect of adding correlation in the calculations for the
cubic thin films in order 
to determine if electron-electron correlation in these structures is sufficient 
to obtain a metal-insulator transition.
Here we use a $U_{\rm eff}=2$~eV, which although larger than that we described 
earlier to more accurately reproduce the PES spectra, does allow us to verify 
that in the absence of insulating behavior, the driving force for the 
MI-transition is not due to intrinsic correlation effects.
Although the numbers we discuss here are particular to a three unit cell thin film 
we note that the general trends are consistent across the series of cubic 
thin films.
In contrast to the LSDA calculations, when a finite Hubbard $U$ is placed on the 
Ru 4$d$ states, we find that the majority $e_g$ states are completely 
unoccupied, and occupation of the majority O 2$p$ states near -2.3~eV is 
enhanced over the minority O 2$p$ states which nearly open a gap in the minority 
spin channael.
An enhancement in the exchange splitting for the Ru $d$ orbitals is also 
observed with $U_{\rm eff}=2$~eV compared to $U_{\rm eff}=0$~eV, while 
the valence bandwidth is reduced.
For the cubic slab with $U_{\rm eff}=2$~eV, the narrowing of the bandwidth 
nearly stabilizes a half-metallic ground state, as the majority Ru $t_{2g}$ bands become completely filled.
Despite these small changes in the occupation of the Ru 4$d$ levels, we do not 
find an insulating ground state in any of the cubic slabs even in the presence 
of strong correlations ($U_{\rm eff}<6$~eV).
Regarding the magnetic moment in these slabs, we find 2.0~$\mu_B$/f.u.\ for 
$U_{\rm eff}=2$~eV, and a corresponding spin polarization at the Fermi level 
of -85.9\%.
These results are consistent with the effects of adding correlation in bulk 
cubic SrRuO$_3$, and because an insulating ground state is not achieved, we 
suggest that neither correlations nor structural confinement 
(from our previous discussion) are sufficient to induce a metal-insulator 
transition.
{\it Orthorhombic LSDA slabs$\quad$}%
We now address films of orthorhombic SrRuO$_3$ which allow the full  
octahedral distortions found in the bulk experimental structure to occur.
Earlier we  showed that the effect of these distortions in the bulk is to 
reduce the $t_{2g}$ valence bandwidth; in this section we examine whether 
these distortions with the addition of a confined geometry in a thin film 
form can stabilize the experimentally observed insulating SrRuO$_3$ ground 
state.
With the relaxed coordinates for bulk $Pbnm$ SrRuO$_3$, we calculate the 
electronic ground state for a three unit cell thick slab separated by 
10~\AA\ of vacuum on each side and SrO termination layers within both the LSDA 
and LSDA+$U$ method ($U_{\rm eff}=2$~eV).\footnote{Complete structural 
relaxation of the three unit cell slab within the orthorhombic symmetry 
did not produce signifigant changes in the electronic or magnetic structure.}
We now discuss the changes in the electronic structure of the orthorhombic thin 
film: In Figure \ref{fig:3uc_slab_doses} we show the (P)DOS for for the three 
unit cell slab with and without correlation.
With the LSDA in the thin film system, the exchange splittings are similar to 
the bulk LSDA orthorhombic calculations, and the character around the Fermi 
level remains a mixture of majority and minority Ru $t_{2g}$.
Energy gaps similar to those found in the bulk are observed in other regions of
the electronic structure with the exception 
that an additional gap opens in the minority $t_{2g}$ states at -2.5~eV, which 
is not observed in the bulk calculation.
We now compare the magnetic properties of the LSDA slab 
calculation to the bulk $Pbnm$ LSDA calculation.
With the LSDA method, we find a magnetic moment of 1.01~$\mu_B$ per Ru 
atom and a spin polarization at the Fermi level of $P_0^{\epsilon_F}=-7.96\%$ 
(compared to a bulk orthorhombic structure where a moment of 0.79~$\mu_B$, 
and spin-polarization of -2.95\%).
Therefore, we find enhanced magnetic properties in the thin film geometry when 
the octahedral tiltings are included. However we still do not find an insulating 
ground state.
{\it Orthorhombic LSDA+$U$ slabs$\quad$}%
Finally we incorporate
correlation into the orthorhombic slab calculations and examine the effect on the 
electronic and magnetic structure.
We have already demonstrated that a $U_{\rm eff}=2$~eV is sufficient to 
establish a half-metallic ground state in the bulk orthorhombic structure; therefore, 
we use this limit to establish whether correlation can drive the insulating 
ground state.
If we do not find a metal-insulator transition even at this large 
Hubbard $U$ value, we can be certain that the effect is not due to correlation.
In general, the shape and weight of the densities of states and exchange 
splittings for the different states remains similar to the 
orthorhombic LSDA slab calculation, however unlike the cubic slabs, the valence 
bandwidth does not noticably narrow.
With the addition of the Hubbard $U$ term in the calculation, the half-metallic 
ground state becomes stable, with 0.70~eV energy gap opening in the majority 
spin states.
This behavior is realized by the minority $t_{2g}$ bands shifting higher in energy 
while the majority bands become completely occupied.
The majority $e_g$ band is also lowered in energy from 1.0~eV in the LSDA slab 
calculation ($U_{\rm eff}=0$~eV) to 0.50~eV, while the minority spin-states move 
0.30~eV higher in energy with $U_{\rm eff}=2$~eV.
Similar energy gaps are observed as in the LSDA slab calculation, with the caveat 
that there is no gap in the majority O 2$p$ states below the Fermi level; this is 
due to a shift of the O 2$p$ states from the Fermi level to lower energy when 
correlation is added.
With a $U_{\rm eff}=2$~eV we find -100\% spin-polarization at the Fermi level 
and a magnetic moment of 2.0~$\mu_B$ per Ru atom.
This effect on the magnetism with increased correlation is consistent with that 
found in the bulk calculations.
In summary, we never find a fully insulating ground state in our 
thin film calculations even in the presence of large correlation effects.
We have also examined the layer-by-layer DOS for each slab (data not shown) 
and have not found an insulating surface layer in any of the calculations.
However, as a result of the 2D confinement in the slabs, we do observe a 
narrowing of the minority $t_{2g}$ bandwidth, and a shift of the Fermi level 
away from the band-center.
Furthermore, with correlations the Fermi level also is seen to cut across the 
band-edge.
These two properties together indicate that SrRuO$_3$ thin films are closer to 
a metal-insulator instability (with regards to the bulk), and consequently 
disorder is more likely to induce electron localization and form an insulating state.
From these results we suggest the following two possibilities regarding the 
experimentally observed metal-insulator transition: 
(1) either the transition in SrRuO$_3$ thin films is not an 
intrinsic property of the system, but rather extrinsic and possibly due to 
surface roughness or defects from film deposition combined with the band 
narrowing from confinement and correlation; or
(2) that SrRuO$_3$ thin films must be treated with more exotic 
electronic structure methods.
It is worth mentioning that PES experiments\cite{Kim_et_al:2005} of 
SrRuO$_3$ films grown on SrTiO$_3$ substrates found a strong 
sensitivity of the $t_{2g}$ spectral intensity and weight early in the 
deposition process (less than eight monolayers) with the intensity of 
the Ru 4$d$ states becoming strongly enhanced above 15 monolayers.
It was found that the film growth proceeds with a step terrace mechanism with 
minor atomic diffusion at less than five monolayers and followed by 3D island 
growth.\cite{Toyota_et_al:2006}
The disordered growth process should also reduce the stability of the 
ferromagnetic order, and due to poor percolation pathways, could contribute to 
the observed MI-transition concomitant with ferromagnetic 
ordering at less than five monolayers.
The disorder at the surface has also recently been compared to that at 
the interface with the substrate (in this case SrTiO$_3$) with {\it in situ} 
PES techniques, and it was found that the sharp $t_{2g}$ peak at the 
Fermi level is greatly suppressed at the surface, while it persists at the 
interface.\cite{Kumigashira_Fujimori_et_al:2008}
The decrease in itineracy due to the suppressed DOS at the Fermi level 
was also verified with surface and interface conductivity 
experiments.
Since our calculations do not include any disordered surface 
configurations or non-stoichiometry, future first-principles calculations 
could clarify these competing interactions.
We have however shown that neither strong correlations nor octahedral 
distortions, nor their combination are sufficient to reproduce the 
experimentally observed ultra-thin film metal-insulator transition.
\section{Conclusions}
We have examined the effects of structural distortions and correlation effects 
on the electronic and magnetic properties of SrRuO$_3$ with first-principles 
calculations.
We find that by including weak strong-correlations with an effective 
Hubbard $U$ of 0.6~eV or correction of the self-interaction error 
gives good agreement for bulk orthorhombic SrRuO$_3$ with the 
experimental spectroscopic data.
The addition of the octahedral distortions leads to a narrowing of the 
majority spin Ru $t_{2g}$ and O 2$p$ states; however the exchange 
splitting is small with respect to these bandwidths and consequently a 
fully insulating ground state is not obtained.
A half-metallic ground state was shown to be stable by including moderate 
electron-electron correlation effects $U > 2$~eV, which we note has not  
been observed experimentally.
The behavior of thin films was also examined in both cubic and 
orthorhombic unsupported films within the conventional LSDA approach and with 
weak correlations included.
In neither case was the experimentally observed metal-insulator transition obtained.
Since the electronic structures of surfaces are very sensitive to atomic 
reconstructions, we suggest that the experimentally observed metal-insulator 
transition could be a consequence of extrinsic defects or an atomically disordered  
surface configuration.

\begin{acknowledgments}
We thank A.\ Fujimori for helpful discussions and bringing our attention to 
correlation characteristics in the spectroscopic data.
The authors also thank A.\ Zayak, J.\ Neaton and W.\ Siemon for 
useful discussions and J.\ Okamoto and H.\ Kumigashira for providing us 
permission and use of the experimental PES data.
This work was supported by the NSF under the grant NIRT 0609377 (NAS), by the 
SFI under the grant 07/IN.1/I945 (NC, SS) and by Seagate.
JMR acknowledges support through a NDSEG fellowship sponsored by the DoD.
Portions of this work made use of MRL Central Facilities supported by the 
MRSEC Program of the National Science Foundation under award No.\ 
DMR05-20415 and the CNSI Computer Facilities at UC Santa Barbara under 
NSF award No.\ CHE-0321368. 
Additional computational resources have been provided by the HEA 
IITAC project managed by the Trinity Center for High Performance 
Computing and by ICHEC.
\end{acknowledgments}


\end{document}